\documentclass[longauth]{aa1} % for the long lists of affiliations 
\usepackage[varg]{txfonts}
\usepackage{graphicx}
\usepackage{color}
\usepackage{lineno}
\usepackage{url}

\usepackage{txfonts}

\begin{document}

   \title{First Multi-wavelength Campaign on the Gamma-ray-loud Active Galaxy IC\,310}

   \author{
M.~L.~Ahnen\inst{1} \and
S.~Ansoldi\inst{2,}\inst{25} \and
L.~A.~Antonelli\inst{3} \and
C.~Arcaro\inst{4} \and
A.~Babi\'c\inst{5} \and
B.~Banerjee\inst{6} \and
P.~Bangale\inst{7} \and
U.~Barres de Almeida\inst{7,}\inst{26} \and
J.~A.~Barrio\inst{8} \and
J.~Becerra Gonz\'alez\inst{9,}\inst{10,}\inst{27}
% ...
 }
 
 \institute{ 
    ETH Zurich, CH-8093 Zurich, Switzerland
  \and Universit\`a di Udine, and INFN Trieste, I-33100 Udine, Italy
  \and INAF National Institute for Astrophysics, I-00136 Rome, Italy
  \and Universit\`a di Padova and INFN, I-35131 Padova, Italy
  \and Croatian MAGIC Consortium, Rudjer Boskovic Institute, University of Rijeka, University of Split - FESB, University of Zagreb - FER, University of Osijek,Croatia
  \and Saha Institute of Nuclear Physics, 1/AF Bidhannagar, Salt Lake, Sector-1, Kolkata 700064, India
  \and Max-Planck-Institut f\"ur Physik, D-80805 M\"unchen, Germany
  \and Universidad Complutense, E-28040 Madrid, Spain
  \and Inst. de Astrof\'isica de Canarias, E-38200 La Laguna, Tenerife, Spain
  \and Universidad de La Laguna, Dpto. Astrof\'isica, E-38206 La Laguna, Tenerife, Spain
  \and University of \L\'od\'z, PL-90236 Lodz, Poland
  \and Deutsches Elektronen-Synchrotron (DESY), D-15738 Zeuthen, Germany
  \and Institut de Fisica d'Altes Energies (IFAE), The Barcelona Institute of Science and Technology, Campus UAB, 08193 Bellaterra (Barcelona), Spain
  \and Universit\`a  di Siena, and INFN Pisa, I-53100 Siena, Italy
  \and Universit\"at W\"urzburg, D-97074 W\"urzburg, Germany
  \and Institute for Space Sciences (CSIC/IEEC), E-08193 Barcelona, Spain
  \and Technische Universit\"at Dortmund, D-44221 Dortmund, Germany
  \and Finnish MAGIC Consortium, Tuorla Observatory, University of Turku and Astronomy Division, University of Oulu, Finland
  \and Unitat de F\'isica de les Radiacions, Departament de F\'isica, and CERES-IEEC, Universitat Aut\`onoma de Barcelona, E-08193 Bellaterra, Spain
  \and Universitat de Barcelona, ICC, IEEC-UB, E-08028 Barcelona, Spain
  \and Japanese MAGIC Consortium, ICRR, The University of Tokyo, Department of Physics and Hakubi Center, Kyoto University, Tokai University, The University of Tokushima, Japan
  \and Inst. for Nucl. Research and Nucl. Energy, BG-1784 Sofia, Bulgaria
  \and Universit\`a di Pisa, and INFN Pisa, I-56126 Pisa, Italy
  \and ICREA and Institute for Space Sciences (CSIC/IEEC), E-08193 Barcelona, Spain
  \and also at the Department of Physics of Kyoto University, Japan
  \and now at Centro Brasileiro de Pesquisas F\'isicas (CBPF/MCTI), R. Dr. Xavier Sigaud, 150 - Urca, Rio de Janeiro - RJ, 22290-180, Brazil
  \and now at NASA Goddard Space Flight Center, Greenbelt, MD 20771, USA and Department of Physics and Department of Astronomy, University of Maryland, College Park, MD 20742, USA
% ...
}
 
\date{Received .../ Accepted ... Draft version}

\offprints{Dorit Eisenacher Glawion (dglawion@lsw.uni-heidelberg.de)}

 \abstract
 {The extragalactic very-high-energy gamma-ray sky is rich in blazars. These are jetted active galactic nuclei that are viewed at a small angle to the line-of-sight. Only a handful of objects viewed at a larger angle are known so far to emit above 100\,GeV. Multi-wavelength studies of such objects up to the highest energies provide new insights into the particle and radiation processes of active galactic nuclei.     
 }
 {We report the results from the first multi-wavelength campaign observing the TeV detected nucleus of the active galaxy IC\,310, whose jet is observed at a moderate viewing angle of $10^\circ-20^\circ$. 
  }
  % methods heading (mandatory)}
 {The multi-instrument campaign was conducted between 2012 November and 2013 January, and involved observations with MAGIC, \textit{Fermi}, \textit{INTEGRAL}, \textit{Swift}, OVRO, MOJAVE and EVN. These observations were complemented with archival data from the AllWISE and 2MASS catalogs. A one-zone synchrotron self-Compton model was applied to describe the broad-band spectral energy distribution.  
  }
 {IC\,310 showed an extraordinary TeV flare at the beginning of the campaign, followed by a low, but still detectable TeV flux. Compared to previous measurements in this energy range, the spectral shape was found to be steeper during the low emission state. Simultaneous observations in the soft X-ray band showed an enhanced energy flux state and a harder-when-brighter spectral shape behaviour. No strong correlated flux variability was found in other frequency regimes. The broad-band spectral energy distribution obtained from these observations supports the hypothesis of a double-hump structure.     
  }
 {The harder-when-brighter trend in the X-ray and VHE emission, observed for the first time during this campaign, is consistent with the behaviour expected from a synchrotron self-Compton scenario. The contemporaneous broad-band spectral energy distribution is well described with a one-zone synchrotron self-Compton model using parameters that are comparable to those found for other gamma-ray-emitting misaligned blazars.
  }

               
\authorrunning{M.L. Ahnen et al.}
\titlerunning{First Multi-wavelength Campaign on the Gamma-ray-loud Active Galaxy IC\,310}
   \maketitle
%
%________________________________________________________________

\section{Introduction}

An active galactic nucleus (AGN) emits radiation over a broad band of the electromagnetic spectrum. Radio-loud AGNs form a subclass in which plasma jets are found to be perpendicularly extending away from the central region consisting of an accretion disk and a supermassive black hole (BH). In the very-high-energy (VHE) gamma-ray range ($50\,\mathrm{GeV}\lesssim E\lesssim50\,\mathrm{TeV}$), sixty-six of these objects have been detected so far.\footnote{\url{http://tevcat.uchicago.edu/}} Most of these objects fall into the subcategory of blazars. They are characterized by strong variability in all energy bands and on all time scales. According to the unified scheme for radio-loud AGNs (Urry \& Padovani \cite{urry95}), blazars are believed to be AGNs viewed at a small angle between the jet-axis and the line-of-sight. Hence, a strong Doppler beaming effect is expected to play a major role in the explanation of the observational properties. Only a few of the detected VHE objects belong to the class of radio galaxies or misaligned blazars: Centaurus\,A \cite{aharonian09}, M\,87 (\cite{aharonian03}; \cite{aharonian06}; \cite{acciari08}; \cite{albert08}; \cite{acciari09}), NGC\,1275 (\cite{aleksic12}, \cite{aleksic14c}), IC\,310 (Aleksi{\'c} et al. \cite{aleksic10}, \cite{aleksic14a}, \cite{aleksic14b}), and PKS\,0625$-$354 \cite{dyrda15}. Radio galaxies and misaligned blazars are viewed at a larger angle to the jet-axis; therefore, the Doppler boosting effect is smaller compared to blazars. 

Due to the small Doppler-boosting effect and often measurable viewing angle, various acceleration and radiation models for the high-energy emission of radio-loud AGNs can be well studied for radio galaxies. This investigation requires multi-wavelength (MWL) data of such objects, preferably simultaneous and with good observational coverage due to their variable behaviour. For all these objects, except for IC\,310 and PKS\,0625$-$354, extensive MWL campaigns up to the VHE range have been conducted and reported previously (Abdo et al. \cite{abdo2009b}; \cite{acciari09}; Abdo et al. \cite{abdo2010a}; Aleksi{\'c} et al. \cite{aleksic14c}). 

IC\,310 is located on the outskirts of the Perseus galaxy cluster with a redshift of $z=0.0189$ \cite{bernardi02}. Originally, this object was classified as a head-tail radio galaxy (\cite{ryle68}; \cite{miley80}; \cite{sijbring98}). However, observations in different frequency bands indicated a transitional object (Aleksi{\'c} et al. \cite{aleksic14a}) with a viewing angle of $10^\circ\lesssim\theta\lesssim20^\circ$ (Aleksi{\'c} et al. \cite{aleksic14b}), showing properties similar to a radio galaxy, e.g., extended radio emission on kpc scales, and a blazar, e.g, a one-sided parsec-scale jet (\cite{kadler12}). While weak optical emission lines observed from IC\,310 are typically found in radio galaxies \cite{owen96}, \cite{rector99} identified IC\,310 as possible low-luminosity BL~Lac object. The X-ray emission is mostly point-like as observed with \textit{ROSAT} and \textit{XMM-Newton} (\cite{schwarz92}; \cite{rhee94}; \cite{sato05}) whereas a hint of X-ray halo emission in the direction of the observed kpc radio jet has been reported by Dunn et al. (\cite{dunn10}). In the soft X-ray band, the flux and spectrum vary in a manner typical for blazars (Aleksi{\'c} et al. \cite{aleksic14a}).
In the gamma-ray band, IC 310 was first detected with the \textit{Fermi}-Large Area Telescope (LAT) at energies above 30\,GeV by \cite{neronov10} and with the MAGIC telescopes above 260\,GeV (Aleksi{\'c} et al. \cite{aleksic10}).  

In this paper we present the results from the first MWL campaign, conducted between 2012 and 2013.
The publication is structured as follows: the observations of all participating instruments and the data analysis are described in Sect. 2 from higher to lower frequencies. In Sect. 3, the observational results are presented. The assembled MWL light curve and spectral energy distribution (SED) will be discussed in Sect. 4, followed by the summary and conclusions in Sect. 5.

\section{Multi-wavelength Observations and Data Analysis}

The MWL campaign for IC\,310 in 2012 and 2013 included observations from radio up to the highest energies with space- and ground-based telescopes. Even while the campaign did not aim at observing the source in a high state,
serendipitously a bright TeV flare with minute time-scale variability was detected in 2012 November by MAGIC (Aleksi{\'c} et al. \cite{aleksic14b}). Participating instruments in the radio band were OVRO (single-dish) as well as the very-long-baseline interferometry (VLBI) arrays: European VLBI Network (EVN) and the VLBA through the ``Monitoring Of Jets in Active galactic nuclei with VLBA Experiment'' (MOJAVE) project at cm wavelengths. In the optical/ultraviolet band, measurements were provided by KVA and \textit{Swift}-UVOT. The X-ray regime was covered by \textit{Swift}-XRT and -BAT, and \textit{INTEGRAL}. \textit{Fermi}-LAT and MAGIC permitted the high-energy (HE, $20\,\mathrm{MeV}\lesssim E\lesssim100\,\mathrm{GeV}$) and VHE gamma-ray measurements.\\
In the following section, the observations of IC\,310 and the data analysis is described.

\subsection{Very high energy: The MAGIC telescopes}

MAGIC is a system of two Imaging Air Cherenkov telescopes, both 17\,m in diameter, located on the Canary Island of La Palma, Spain. It covers the electromagnetic spectrum in the VHE range from $50$\,GeV to $50$\,TeV and achieves an  angular resolution of $\sim0.1^\circ$ (Aleksi{\'c} et al. \cite{AleksicSoftwareUpgrade}). 

The observations during the MWL campaign were conducted after the upgrade of the two telescopes in 2011-2012 was completed \cite{AleksicHardwareUpgrade}. On the first night of observation, 2012 November 12-13 (MJD 56243.95-56244.11), during 3.7\,h of observation, MAGIC detected a bright flare as reported in Aleksi{\'c} et al. (\cite{aleksic14b}). 
Further observations until 2013 January 17 (MJD 56309.1) have been carried out as part of the MWL campaign during dark and moon time. Data affected by non-optimal weather conditions were discarded.
Only observations during dark night and with moderate moon light were selected, and the standard analysis (Aleksi{\'c} et al. \cite{AleksicSoftwareUpgrade}) can be applied. After the selection, the data set consists of $\sim39$\,h including the data of the flaring night.
The data cover the zenith distance range of $11^\circ<\mathrm{Zd}<\,56^\circ$. 

The analysis of the data is performed analogously to Aleksi{\'c} et al. (\cite{aleksic14b}) and Aleksi{\'c} et al. (\cite{AleksicSoftwareUpgrade}). 
The image cleaning is performed using the dynamical sum-cleaning algorithm presented in \cite{sitarek13}. The significance of the signal is calculated from Eq.\,17 of \cite{lima83} using four background regions that do not overlap with the emission from NGC\,1275, which is a VHE object $0.6^\circ$ away from IC\,310. Flux and differential upper limits are calculated according to \cite{rolke} using a 95\% confidence level. 
Following Aleksi{\'c} et al. (\cite{AleksicSoftwareUpgrade}), we consider for the spectra the following systematic errors: 11\% for the flux normalization, 15\% for the energy scale and 0.15 for the photon index.
As reported previously in Aleksi{\'c} et al. (\cite{aleksic14a}, \cite{aleksic14b}), the absorption due to the extragalactic background light (EBL) is only marginal for IC\,310. The intrinsic spectra presented in this paper are calculated using the model of \cite{dominguez11}.

\subsection{High energy: \textit{Fermi}-LAT}

\textit{Fermi} was launched in 2008 June and since 2008 August 5, it is operated primarily in sky survey mode, scanning the entire sky every three hours  \cite{atwood}. The {\it Fermi}-LAT is a pair-conversion telescope sensitive to photons between 20~MeV and several hundred GeV \cite{ackermann12}.

We calculated spectra using the \textit{Fermi} Science Tools (v10r0p5)\footnote{Available online at \url{http://fermi.gsfc.nasa.gov/ssc/data/analysis/software/}.} for the time period 2012 November 01 (MJD\,56232) to 2013 January 31 (MJD\,56323). We used the Pass 8 data, the recommended {\it P8R2\_SOURCE\_V6} instrumental response functions, the isotropic diffuse background template {\it iso\_P8R2\_SOURCE\_V6\_v06} and the Galactic emission model {\it gll\_iem\_v06} \cite{acero2016}\footnote{\url{http://fermi.gsfc.nasa.gov/ssc/data/access/lat/BackgroundModels.html}}, a region of interest (ROI) with a radius of 10$^\circ$ and an energy range of 1\,GeV - 300\,GeV.
We ran the unbinned likelihood analysis with a 90$^\circ$ zenith angle cut to reduce contamination from the Earth limb.
As an input for the likelihood analysis we used the 3FGL model \cite{acero2015}. For sources within the ROI the photon index and prefactor (and equivalent for other models) were left free. Spectral upper limits are calculated with a limiting Test Statistic (TS) of 25 \cite{mattox1996}, and taking into account sources in the spectral model of a radius of 20$^\circ$. The predicted number of counts is low ($\sim$10).
Because NGC\,1275 appears to be very bright and close to IC\,310 with an offset of $0.6^\circ$, the results reported here are calculated for energies higher than 1\,GeV to mitigate the effects of a larger point-spread function at lower energies.\footnote{The plot of the point-spread function can be found online at \url{https://www.slac.stanford.edu/exp/glast/groups/canda/lat_Performance.htm}.}

\subsection{X-ray: \textit{INTEGRAL}, \textit{Swift}-BAT/XRT }

The \textit{International Gamma-Ray Astrophysics Laboratory} (\textit{INTEGRAL}) satellite is in operation since late 2002 \cite{winkler03}. It is equipped with several instruments in the hard X-ray to soft gamma-ray range: the high resolution spectrometer SPI at 20\,keV--8\,MeV \cite{vedrenne03}, and two high angular resolution gamma-ray imagers, called IBIS, which operate at 15--1000\,keV and 0.175--10.0\,MeV \cite{Ubertini03}. 

The IBIS data are extracted using the Offline Science Analysis tool OSA, version 10.1. All data between 2012 August and 2013 February are taken into account, where IC\,310 was located within $14^\circ$ from the pointing center, resulting in 451 science windows. These data are filtered for the energy range between 20\,keV and 200\,keV. No significant signal is detected from these data and upper limits are derived. SPI data are not used for this publication.

The \textit{Swift} satellite was launched in late 2004 \cite{gehrels04}. \textit{Swift} provides measurements with telescopes covering the optical and X-ray (soft and hard) ranges.
Continuous observations in the hard X-ray range (15--150\,keV), mainly for detecting gamma-ray bursts, are provided by the Burst Alert Telescope (BAT). The X-Ray Telescope (XRT) operates in the soft X-ray regime from 0.2 to 10\,keV \cite{burrows}. 

We extract a spectrum in the energy band of 20--100\,keV from the 104-month \textit{Swift}-BAT survey maps and fit the spectrum with a simple power law with fixed normalization. A fitting statistic for Poisson-distributed source count rates and Gaussian-distributed background count rates as recommended in the XSPEC statistics appendix is used.\footnote{\url{https://heasarc.gsfc.nasa.gov/xanadu/xspec/manual/XSappendixStatistics.html}}
The calculations of the flux, photon index, and the corresponding 90\% uncertainties are
implemented using Monte Carlo simulations of the spectrum.

Furthermore, observations of IC\,310 with the \textit{Swift}-XRT with a total exposure of 45.8\,ks were performed in 2012 November and December. The results presented here are compared to earlier observations of 13.2\,ks taken in January of the same year. The XRT data were reduced with standard methods, using the most recent software packages (HEASOFT 6.15.1 3) and calibration databases. Spectra were grouped to a minimum signal-to-noise ratio of 5 to ensure the validity of $\chi^2$ statistics. For the broad-band SED we applied another re-binning in order to increase the significance of individual points. Spectral fitting was performed with ISIS 1.6.2 \cite{houck00}. We fitted the 0.5--10 keV energy band with an absorbed power-law model, which yielded a good fit probability. X-ray data were de-absorbed using abundances from \cite{wilms00} and cross sections from \cite{verner96}. As previous \textit{Chandra} observations (Aleksi{\'c} et al. \cite{aleksic14a}) revealed a $N_\mathrm{H}$ significantly above the 
Galactic $N_\mathrm{H}$ value of $0.12\times10^{22}$cm$^{-2}$ for IC\,310 (Kalberla et al. \cite{kalberla10}) we left $N_\mathrm{H}$ free in the fit.

\subsection{Ultraviolet and Optical: \textit{Swift}-UVOT, KVA}

In addition to the X-ray instruments, \textit{Swift} is equipped with the UltraViolet/Optical Telescope (UVOT) providing observations in the ultraviolet (UV) and optical ranges simultaneous with the XRT \cite{gehrels04}. The telescope is equipped with the following filters: V (547\,nm), B (439\,nm), U (347\,nm), UVW1 (260\,nm), UVM2 (225\,nm), and UVW2 (193\,nm). Swift-UVOT data were extracted following standard methods.\footnote{\url{http://swift.gsfc.nasa.gov/analysis/UVOT_swguide_v2_2.pdf}}

The Kungliga Vetenskaps Akademien (KVA) telescopes are
located at the Observatorio del Roque de los Muchachos on the
island of La Palma, Spain, and are operated by the Tuorla observatory.\footnote{\url{http://users.utu.fi/kani/1m}}
They consist of two optical telescopes with mirror diameters of 60\,cm and 35\,cm.
Filters in the R-band (640 nm), B-band (550 nm), and V-band
(440 nm) are available. The photometric observations were conducted with the
R-band filter and the 35\,cm telescope during the MAGIC observations. The data were analyzed using a standard semi-automatic pipeline. The brightness of the source is measured using differential photometry with a standard aperture radius of 5.0\(^{\prime\prime}\). The host galaxy emission is expected to be constant. Any AGN variability would affect the light curve.

Optical, infrared, and ultraviolet data were de-reddened using the same absorbing columns obtained from the \textit{Swift}-XRT data (Nowak et al. 2012, and references therein).

\subsection{Radio: OVRO, EVN, VLBA/MOJAVE}

The 40\,m Owens Valley Radio Observatory (OVRO, California, USA) telescope provides radio data for a list of AGNs at 15\,GHz nearly twice per week since 2008.\footnote{\url{http://www.astro.caltech.edu/ovroblazars/}} Details on the observing strategy and the calibration procedures are summarized in \cite{richards11}.  
The data presented in this paper cover the time range from 2012 October 31 (MJD\,56231) to 2012 December 22 (MJD\,56283).    

The European VLBI Network is a consortium of several radio-astronomical institutes and telescopes from Europe, Asia, and South Africa.\footnote{\url{http://www.evlbi.org/}} Due to the large collection area of its telescopes, the EVN provides excellent sensitivity to weak emission. For IC\,310, observations in October and November 2012 at the frequencies 1.7, 5.0, 8.4, and 22.2\,GHz were carried out. 

The MOJAVE project is a long-term VLBI monitoring program 
at 15\,GHz conducted with the Very Long Baseline Array (VLBA) as a continuation of the VLBA 2\,cm survey, e.g., \cite{lister2016}.\footnote{\url{https://science.nrao.edu/facilities/vlba}} The array consists of ten identical 25\,m (in diameter) antennas with a baseline up to 8000\,km.  IC\,310 was included in the target list of MOJAVE in early 2012.

A description of the analysis of the EVN data can be partially found in Aleksi{\'c} et al. (\cite{aleksic14b}). More information on the data analysis procedure of the MOJAVE and OVRO data and an extended analysis of these data will be published in a separate paper (Schulz et al. in prep.), which will include additional observations made over a longer period of time that will allow the investigation of possible changes of the jet structure. Only the results of the OVRO observations during the campaign and EVN and MOJAVE flux density measurements will be presented here because they are relevant for the study of the multi-band light curve and the SED. 

\subsection{Additional data}

We also consider historical data of IC\,310 for the SED from the \textit{Wide-Field Infrared Survey Explorer} (\textit{WISE}) and the Two Micron All Sky Survey (2MASS). The AllWISE Source Catalog \cite{wright10} covers a time range from 2010 January to 2010 November, while the 2MASS catalog \cite{skrutskie06} covers from 1997 June to 2001 February.
The data are de-reddened using the absorption column derived from spectral fits to the X-ray observations. 
%______________________________________________________________
\section{Results}

In this section we present the results from the MWL observation starting with the highest energies.

\subsection{MAGIC results}
%___________________________________________________________

During 2012 November to 2013 January (MJD 56245.0--56309.1), IC\,310 was detected by MAGIC using data (excluding the flare data of MJD 56244.0) over an effective time (observation time minus dead time) $t_\mathrm{eff}=35.3$\,h with a significance of 5.22\,$\sigma$ above 300\,GeV. This low significance measured over rather a long time span already indicates a low flux state after the flare. Here we have excluded the flare data from the data selection to test the detection of the object outside of the flaring state and because a large part of the observations used for the MWL SED, e.g., in X-rays, was performed not simultaneous to the TeV flare. 

\begin{figure}
    \centering
       \includegraphics[width=9cm]{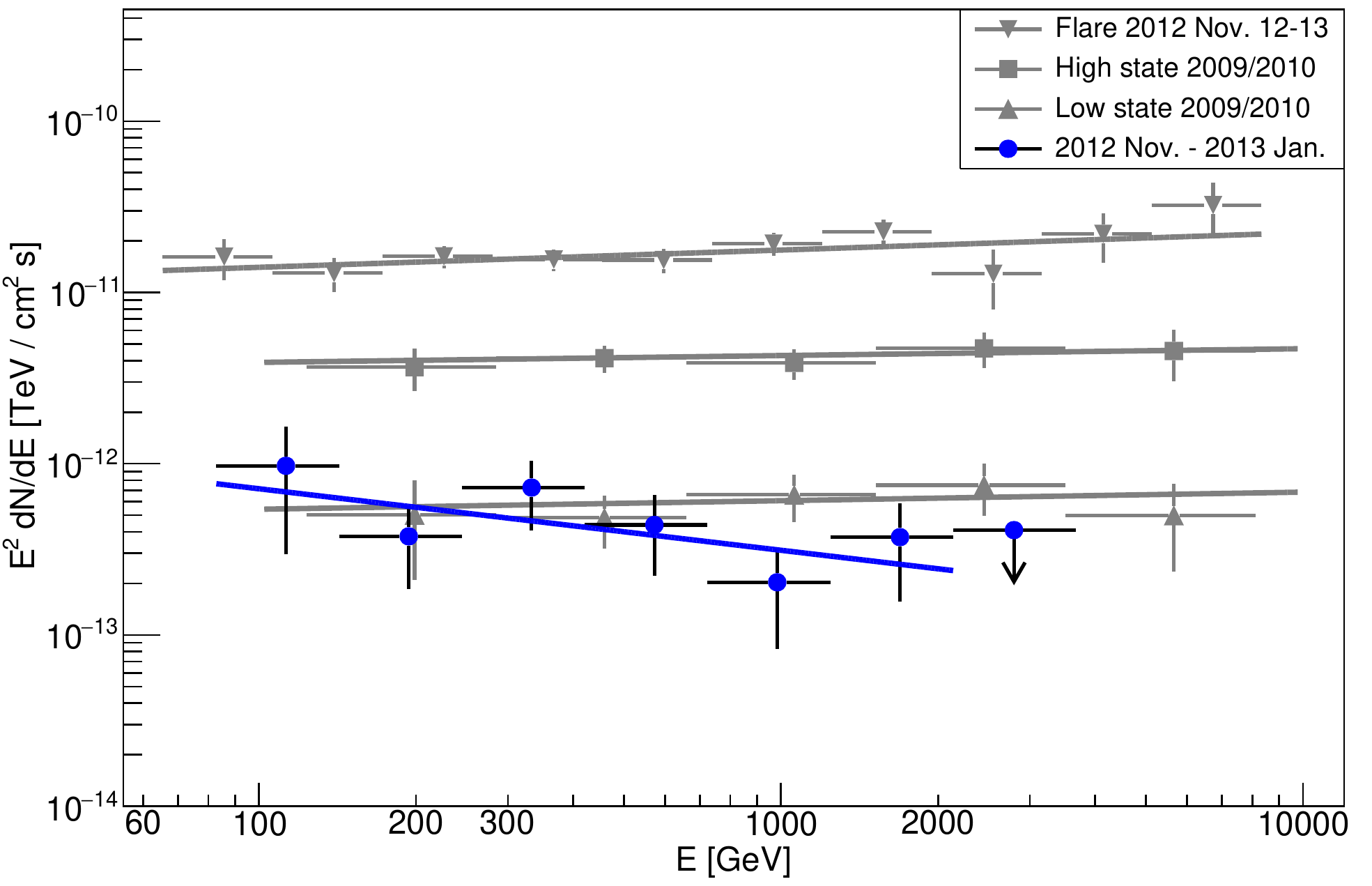}
       \caption{Measured spectra in the VHE band during different flux states obtained with MAGIC. The blue data points as well as the blue line show the resulting spectra from observations during 2012 November and 2013 January (excluding the large VHE flare from 2012 Nov. 12-13). Spectral results from previous publications (Aleksi{\'c} et al. \cite{aleksic14a}, \cite{aleksic14b}) are shown in gray for comparison.}
               \label{VHESpectra1}%
\end{figure}

The energy spectrum $\left(\mathrm{d}N/\mathrm{d}E\right)$ between 82 GeV and 2.1 TeV during the campaign from 2012 November to 2013 January (Fig.~\ref{VHESpectra1}), excluding the flare data, can be described with the power-law function:  
\begin{equation}
 \frac{\mathrm{d}N}{\mathrm{d}E}=f_0\times\left(\frac{E}{1\mathrm{TeV}}\right)^{-\Gamma}\left[\frac{10^{-12}}{\mathrm{cm}^{2}\mathrm{s}\,\mathrm{TeV}}\right],
\end{equation}
with a flux normalization $f_0=(3.12\pm0.91_{\mathrm{stat}}\pm0.34_{\mathrm{syst}})\times10^{-13}\,\mathrm{TeV}^{-1}\,\mathrm{cm}^{-2}\,\mathrm{s}^{-1}$\, at 1\,TeV and a photon spectral index of $\Gamma=(2.36\pm0.30_{\mathrm{stat}}\pm0.15_{\mathrm{syst}})$. The normalization energy of 1\,TeV was fixed and selected for easier comparison with previous measurements.
As simultaneous \textit{Swift}-XRT and UVOT data are only available for the 2012 November and December observations, we also calculated a MAGIC spectrum for the time period 2012 November to December in the same energy range.  
The resulting spectrum agrees, within the errors, with the spectrum calculated for the entire campaign. Thus, in the subsequent sections we only show and discuss the spectrum for the entire campaign.

\begin{table}
\small
\caption{Division of the data of the flare in the night of 2012-11-12 to 2012-11-13 according to different flux states.}            
\label{table:DivisionFlareTimes}      
\centering                         
\begin{tabular}{c c c c c c}        
\hline\hline
period &time start &time stop &  MJD start& MJD stop	\\
       & hh:mm:ss & hh:mm:ss &        & 		\\
\hline
Ia & 22:50:49 & 23:10:35 &56243.951958	&56243.965686\\ 
Ib & 23:55:20 & 01:16:14 &56243.996756	&56244.052941\\ 
II & 23:11:21 & 23:51:18 &56243.966217	&56243.993958\\ 
III & 01:16:52 & 01:36:36 &56244.053380	&56244.067083\\ 
IV & 01:37:20 & 01:56:58 &56244.067593	&56244.081231\\ 
V & 01:57:36 & 02:33:05 &56244.081669	&56244.106308\\ 
\hline      
\end{tabular}
\end{table}

\begin{figure}
    \centering
       \includegraphics[width=9cm]{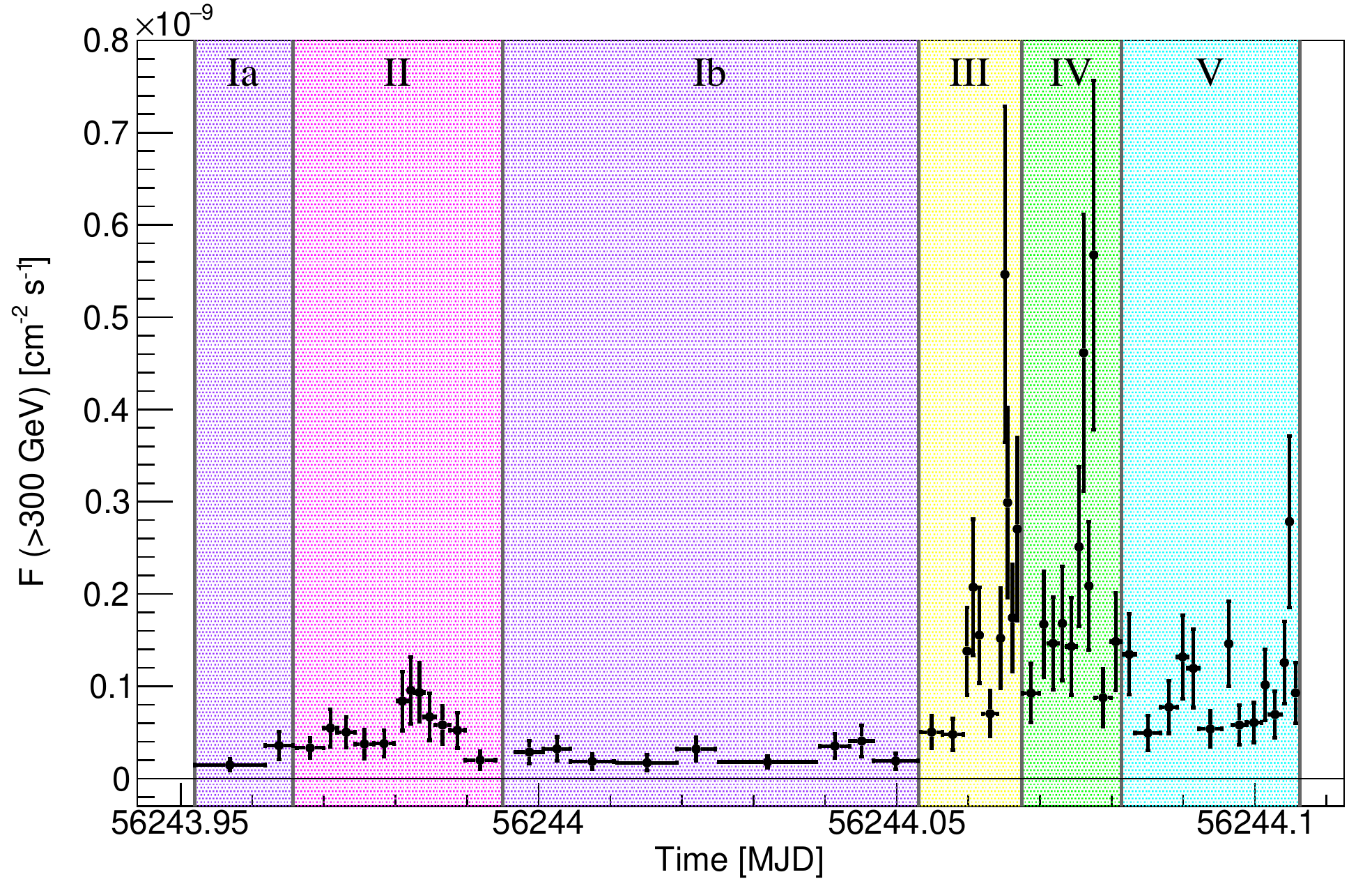}
       \caption{Division of the MAGIC data taken during the night of 2012 November 12-13 into data sub-sets according to different flux states. The light curve data points are taken from Aleksi{\'c} et al. (\cite{aleksic14b}). Vertical lines and colored boxes indicate the boundaries for the different flux periods used for the investigation of the spectral variability. The time intervals related to the above-mentioned periods are reported in Table~\ref{table:DivisionFlareTimes}.}
               \label{MAGIC_lightcurve_flare_lines}
\end{figure}

\begin{figure}
    \centering
       \includegraphics[width=9cm]{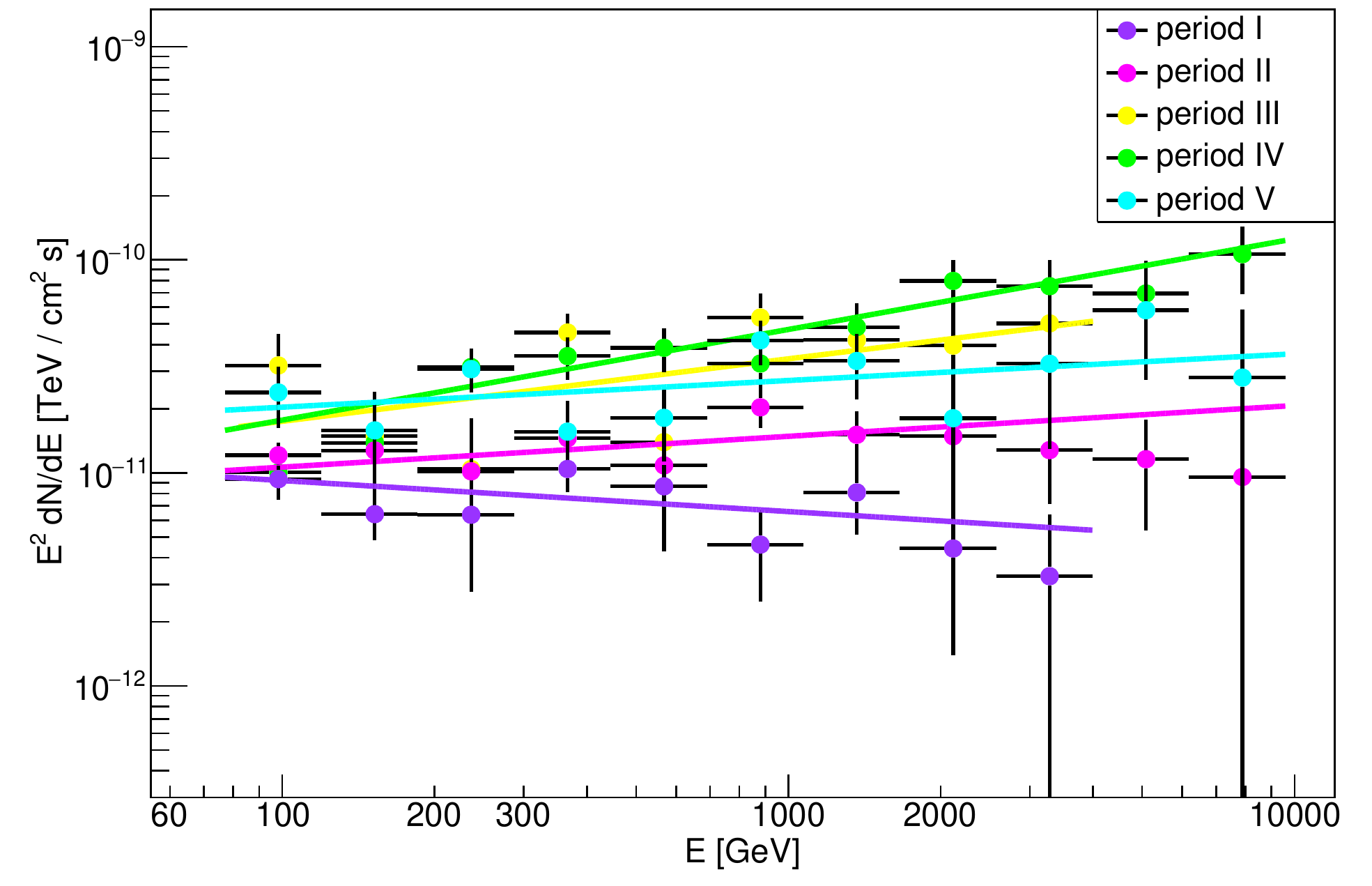}
       \caption{Measured spectral energy distributions from different time periods during the night of the flare 2012 November 12-13 obtained by MAGIC. The time periods are given in Table~\ref{table:DivisionFlareTimes} and indicated with lines in Fig.~\ref{MAGIC_lightcurve_flare_lines}. The same color scheme as in Fig.~\ref{MAGIC_lightcurve_flare_lines} is used.}
               \label{VHESpectra2}%
\end{figure}

\begin{table*}
\caption{Results of power-law spectra measured with MAGIC during different periods. }             
\label{table:VHESpecta}     
\centering                          
\begin{tabular}{l c c c c c}        
\hline\hline
state    & comment  & energy &$f_{0}\pm f_{\mathrm{stat}}\pm f_{\mathrm{syst}}$ &$\Gamma\pm\Gamma_{\mathrm{stat}}\pm\Gamma_{\mathrm{syst}}$ & References\\
         &  & range [TeV] &$\times10^{-12}[\mathrm{TeV}^{-1}\,\mathrm{cm}^{-2}\,\mathrm{s}^{-1}]$& &\\
\hline
high state 2009/2010&observed & 0.12-8.1 &$4.28\pm0.21\pm0.73$     &$1.96\pm0.10\pm0.20$ & Aleksi{\'c} et al. (\cite{aleksic14a})\\ 
low state 2009/2010&observed & 0.12-8.1 &$0.608\pm0.037\pm0.11$   &$1.95\pm0.12\pm0.20$ & Aleksi{\'c} et al. (\cite{aleksic14a})\\
flare Nov 2012&observed  & 0.07-8.3& $17.7\pm0.9\pm2.1$ & $1.90\pm0.04\pm0.15$ & Aleksi{\'c} et al. (\cite{aleksic14b})\\ 
\hline
flare period Nov 2012  I	&observed  &0.07-4.0	&$6.4\pm0.8\pm0.7$	&$2.15\pm0.10\pm0.15$ & this work\\ 
flare period Nov 2012  II	&observed  &0.07-9.5	&$13.5\pm0.9\pm1.5$	&$1.96\pm0.06\pm0.15$ & this work\\ 
flare period Nov 2012  III &observed  &0.07-4.0	&$37.5\pm4.2\pm4.1$	&$1.63\pm0.11\pm0.15$ & this work\\ 
flare period Nov 2012  IV	&observed  &0.07-9.5	&$44.2\pm2.1\pm4.9$	&$1.51\pm0.06\pm0.15$ & this work\\ 
flare period Nov 2012  V	&observed  &0.07-9.5	&$27.5\pm2.3\pm3.0$	&$1.85\pm0.08\pm0.15$ & this work\\ 
\hline
Nov 2012 - Jan 2013&observed &0.08-2.1 	&$0.31\pm0.09\pm0.03$	& $2.36\pm0.30\pm0.15$& this work\\ 
Nov 2012 - Jan 2013&intrinsic &0.08-2.1 &$0.37\pm0.11\pm0.04$  & $2.26\pm0.30\pm0.15$& this work\\
\hline       
\end{tabular}
\end{table*}

We further investigate the change in the spectrum at different flux states by comparing our results with previous measurements, as well as by studying spectra of different flux states during the flare on 2012 November 12-13. In Aleksi{\'c} et al. (\cite{aleksic14a}) and Aleksi{\'c} et al. (\cite{aleksic14b}), no significant spectral variability was reported. Here, we present individual spectra during the flare. The division of the data of the flare according to different flux states is given in Table~\ref{table:DivisionFlareTimes} and shown in Fig.~\ref{MAGIC_lightcurve_flare_lines}. Besides the separations based on the fluxes, the definitions of the intervals are also determined by the duration of observing runs of typically 20 minutes. The resulting spectra are shown in Fig.~\ref{VHESpectra2} and their parameters are listed in Table~\ref{table:VHESpecta}.

\begin{figure}
    \centering
       \includegraphics[width=9cm]{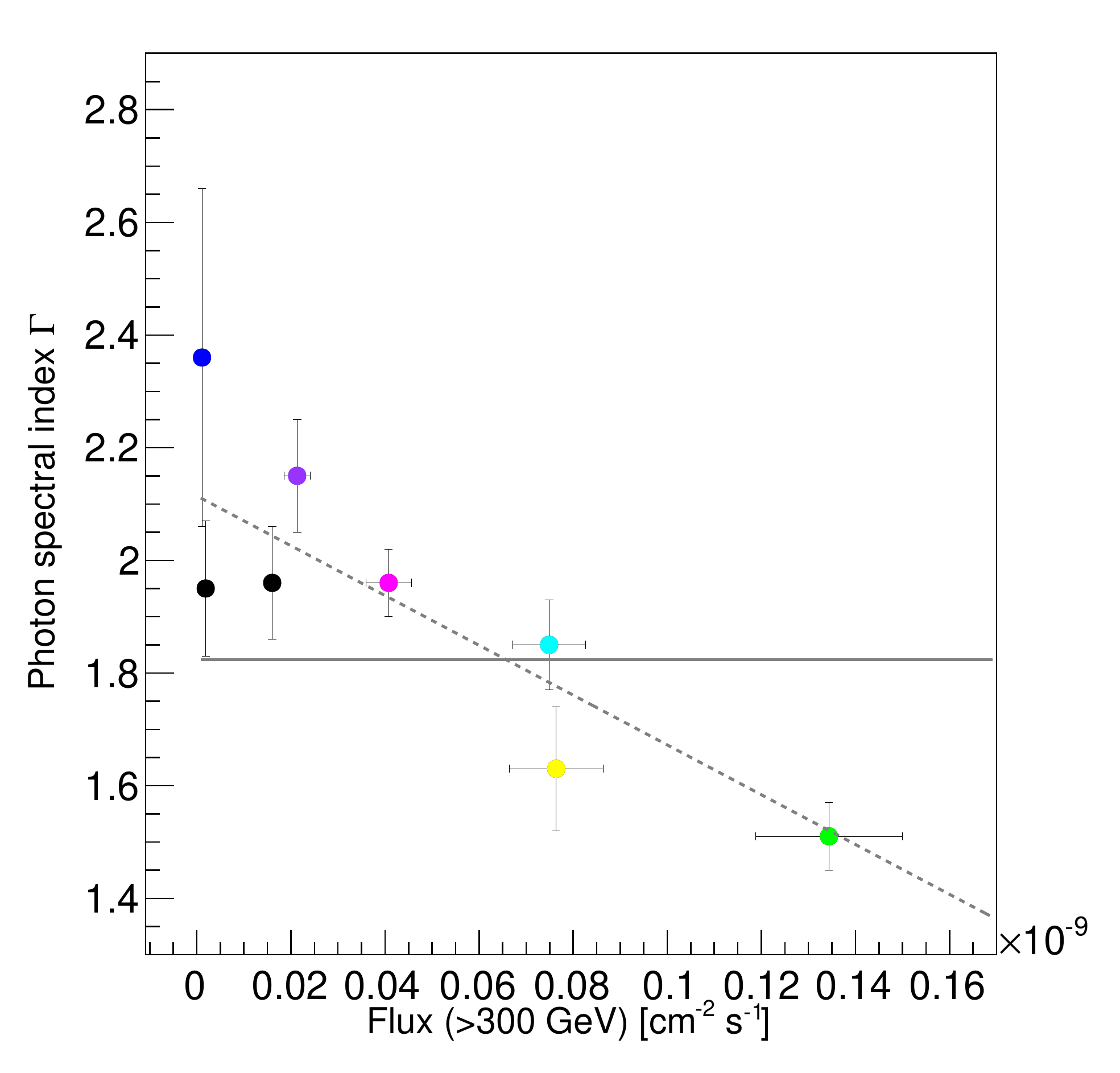}
       \caption{Power-law index of VHE spectra as a function of the VHE flux above 300\,GeV measured by MAGIC. The photon index values, which are not corrected for EBL absorption, are listed in Table~\ref{table:VHESpecta}.
       The color scheme is the same as the one used in Fig.~\ref{MAGIC_lightcurve_flare_lines} and in Fig.~\ref{VHESpectra2}. In addition, the result for the multi-wavelength campaign is shown in blue, and the historical measurements from Aleksi{\'c} et al. (\cite{aleksic14a}) are shown in black. The solid and dashed lines represent a constant and a linear fit, respectively.  }
               \label{VHESpectraCom}
\end{figure}

In Fig.~\ref{VHESpectraCom}, we present for all temporally non-overlapping data the photon spectral index versus the integrated flux (overlapping data would include a bias so the averaged flare spectrum on 2012 November 12 - 13 is not included). All integrated fluxes represent mean fluxes and have been obtained by fitting all light curves above 300\,GeV with a constant line, both the light curves from previous measurements (Aleksi{\'c} et al. \cite{aleksic14a}) and from the data presented here.
A constant fit to the data points in Fig.~\ref{VHESpectraCom} reveals a $\chi^2/\mathrm{d.o.f}$ of 52.5/7, thus a low probability for being constant of $4.6\times10^{-9}$. A higher probability can be obtained with a linear function which yields a $\chi^2/\mathrm{d.o.f}$ of 7.0/6 and a probability of $0.32$.
  
\begin{table}
\small
\caption{MAGIC gamma-ray flux measurements from single observations, as well as integrated over months. The dates given in the first column correspond to the day following the observation night. The upper limits were computed with a 95\% confidence level.}        
\label{table:1}     
\centering                        
\begin{tabular}{c c c c}       
\hline\hline 
   used data             &MJD start     & $t_{\mathrm{eff}}$ & $F_{\mathrm{E}>300\,\mathrm{GeV}}$ \\    
  & & [h] & [$10^{-12}$\,cm$^{-2}$\,s$^{-1}$] \\
\hline 
   all data  (flare excl.)   	& ...                & 35.3         & $1.59\pm0.29$                  \\
\hline
   2012 Nov. (flare excl.) 	& ...		 & 17.9		 & $1.69\pm0.42$\\
   2012 Dec.			&...		 & 7.1		 &$1.88\pm0.61$		\\
   2013 Jan.			&...		 & 10.4		 &$(0.61\pm0.54)$ $< 2.09$	\\
\hline 
2012-11-13 & 56243.95 & 3.51 & $61.82\pm2.92$\\
2012-11-14 & 56245.00 & 2.37 & $< 3.03$\\
2012-11-16 & 56246.94 & 3.47 & $< 1.74$ \\
2012-11-18 & 56249.11 & 2.39 & $< 6.72$ \\
2012-11-19 & 56249.99 & 4.41 & $< 4.70$  \\
2012-11-21 & 56252.06 & 1.30 & $< 4.52$ \\
2012-11-22 & 56253.09 & 1.96 & $< 4.81$ \\
2012-11-23 & 56254.13 & 1.30 & $6.43\pm1.87$ \\
2012-11-24 & 56255.17 & 0.65 & $< 10.26$ \\
2012-12-15 & 56275.93 & 0.92 & $< 8.05$\\
2012-12-16 & 56276.82 & 0.65 & $< 2.83$ \\
2012-12-17 & 56277.83 & 2.01 & $< 5.29$ \\
2012-12-18 & 56278.83 & 3.50 & $3.26\pm0.96$ \\
2013-01-10 & 56301.89 & 0.62 & $5.71\pm2.37$ \\
2013-01-11 & 56302.88 & 0.65 & $< 3.12$ \\
2013-01-13 & 56304.87 & 0.80 & $< 9.21$ \\
2013-01-14 & 56305.82 & 0.98 & $< 6.21$ \\
2013-01-15 & 56306.83 & 1.50 & $< 3.95$ \\
2013-01-16 & 56307.83 & 2.50 & $2.22\pm1.07$  \\
2013-01-17 & 56308.82 & 3.34 & $< 1.35$\\
\hline                                   
\end{tabular} 
\end{table}

The mean integrated flux over the entire period is measured to be $F_{\mathrm{mean}}=(1.59\pm0.29)\times10^{-12}$\,cm$^{-2}$\,s$^{-1}$ above 300\,GeV when excluding the data from the 2012 November flare. This is $\sim40$ times lower than the mean integrated flux of $(6.08\pm0.29)\,\times10^{-11}$\,cm$^{-2}$\,s$^{-1}$ reported for the 2012 November flare in Aleksi{\'c} et al. (\cite{aleksic14b}).  Due to the faint emission, a monthly light curve is computed.  The light curve is calculated assuming a simple power-law distribution with a photon index of $\Gamma=2.4$. For the data during the flare, a different photon index, $\Gamma=2.0$, is used. The monthly calculated light curve is shown in Fig.~\ref{MWL_lightcurve}. Fitting this light curve with a constant reveals a flux of $F_{\mathrm{const.}}=(1.42\pm0.29)\times10^{-12}$\,cm$^{-2}$s$^{-1}$
with a $\chi^2$/d.o.f. of 3.2/2 (probability of 0.20). Thus, no significant variability of the flux is found from month to month. 
Results from individual days are given in Table~\ref{table:1}.

\subsection{Fermi-LAT results}

In the time range from 2012 November 01 (MJD\,56232) to 2013 January 31 (MJD\,56323), IC\,310 could not be detected with \textit{Fermi}-LAT (TS$<25$) over the entire three months period. 
The measured light curve is shown in Fig.~\ref{MWL_lightcurve}.  Only upper limits of the flux and one flux point in 2013 January could be calculated. Thus, no further conclusion can be drawn on the variability behaviour of IC\,310. We also searched for individual gamma-ray event candidates in the \textit{Fermi}-LAT detector during the MWL campaign. Within a circle with a radius of $10.0^\circ$ around IC\,310, only four events could be found above 1\,GeV with a larger likelihood to originate from IC\,310 than from NGC\,1275 and a probability of higher than 50\%. The probability for each photon was calculated using \texttt{gtsrcprob}. They were detected on MJD\,56248.4, 56298.8, 56303.7, and 56318.1, with energies of  4.8, 8.0, 2.4, and 17.2\,GeV, respectively. The arrival times are indicated with blue lines in Fig.~\ref{MWL_lightcurve}.
The calculated 95\% confidence level upper limits for the SED are shown in Fig.~\ref{MWL_SED}. The blue upper limit spectrum covers the full time range of the campaign from 2012 November 01 to 2013 January 31 and an energy range from 1\,GeV to 72\,GeV.

\subsection{X-ray}

  \begin{figure*}
    \centering
    \includegraphics[width=15cm]{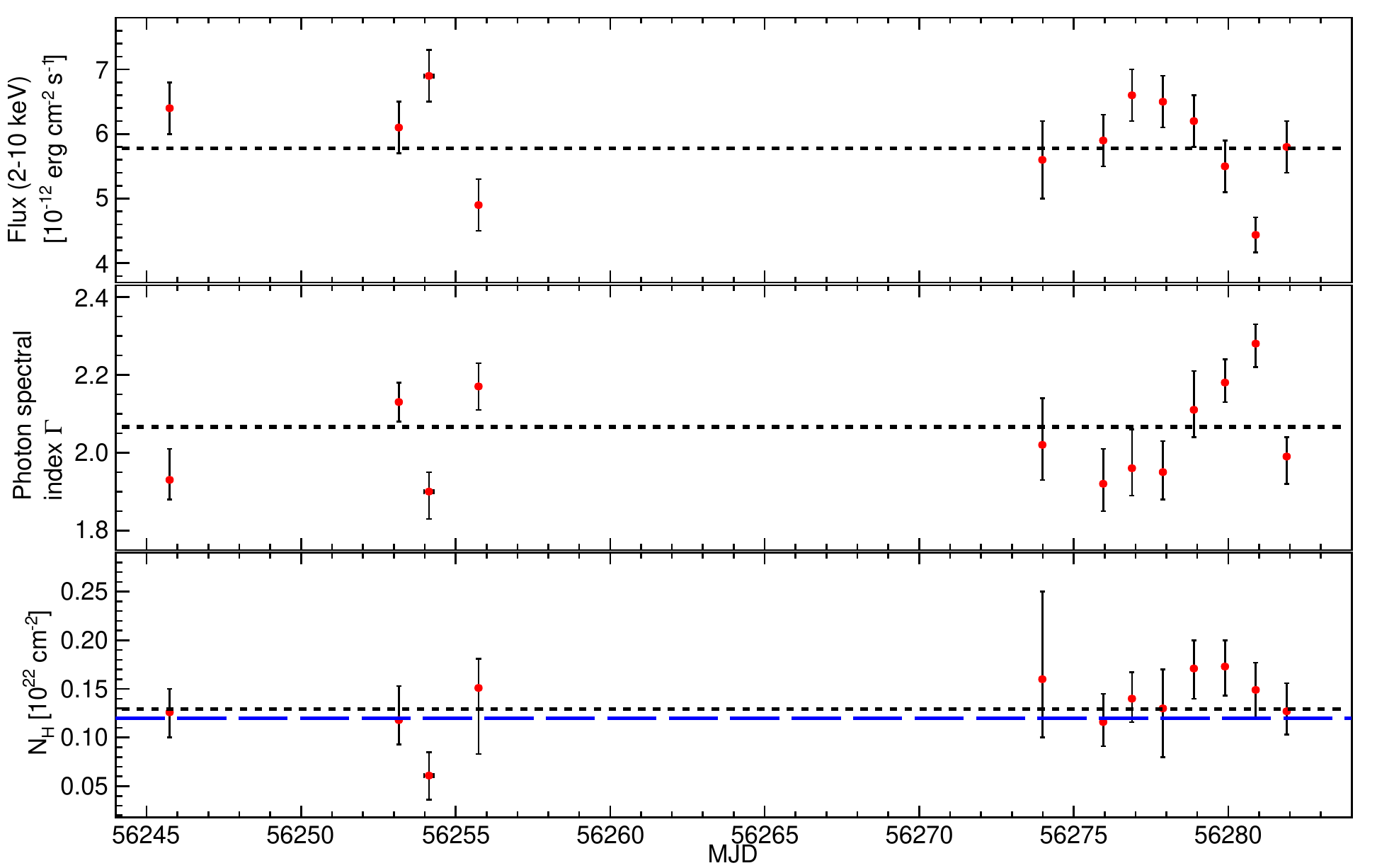}
    \caption{\textit{Swift}-XRT measurements during the campaign. All points have been fitted with a constant (black dashed line). \textit{Top panel:} X-ray light curve in an energy range of 2--10\,keV.
    \textit{Middle panel:} Photon spectral index. 
    \textit{Bottom panel:} Neutral hydrogen column density. For comparison, the Galactic $N_{\mathrm{H}}$ value of $0.12\times10^{22}$\,cm$^{-2}$ for IC\,310 (Kalberla et al.\cite{kalberla10}) is shown as a dashed blue line.}
    \label{XRay1}	
    \end{figure*}

The light curve measured with \textit{Swift}-XRT is shown in Fig.~\ref{XRay1} and in Fig.~\ref{MWL_lightcurve}. An overview of the results - e.g., the flux level in the energy range of 2--10\,keV, the photon index, and $N_\mathrm{H}$ - can be found in Table~\ref{table_LC_swift}. 

The temporal evolution of the flux in the energy range 2--10\,keV, the photon index, and the column density $N_\mathrm{H}$ in 2012 November to December during the campaign is presented in Fig.~\ref{XRay1}. 
The mean energy flux has been measured to be $(0.61\pm0.01)\times10^{-11}$\,erg\,cm$^{-2}$\,s$^{-1}$, which is about five times higher than during previous measurements (Aleksi{\'c} et al. \cite{aleksic14a}) and moderately higher (factor of 1.4) than in 2012 January (see Table~\ref{table_LC_swift}). 
A fit to the light curve with a constant line reveals a probability of $7.3\times10^{-7}$
for a constant energy flux of $(0.58\pm0.01)\times10^{-11}$\,erg\,cm$^{-2}$\,s$^{-1}$ ($\chi^2$/d.o.f.$=49.6/11$). The observation on MJD\,56280.82 deviates by $\sim5\,\sigma$ from this constant energy flux value.

\begin{figure}
    \centering
    \includegraphics[width=9cm]{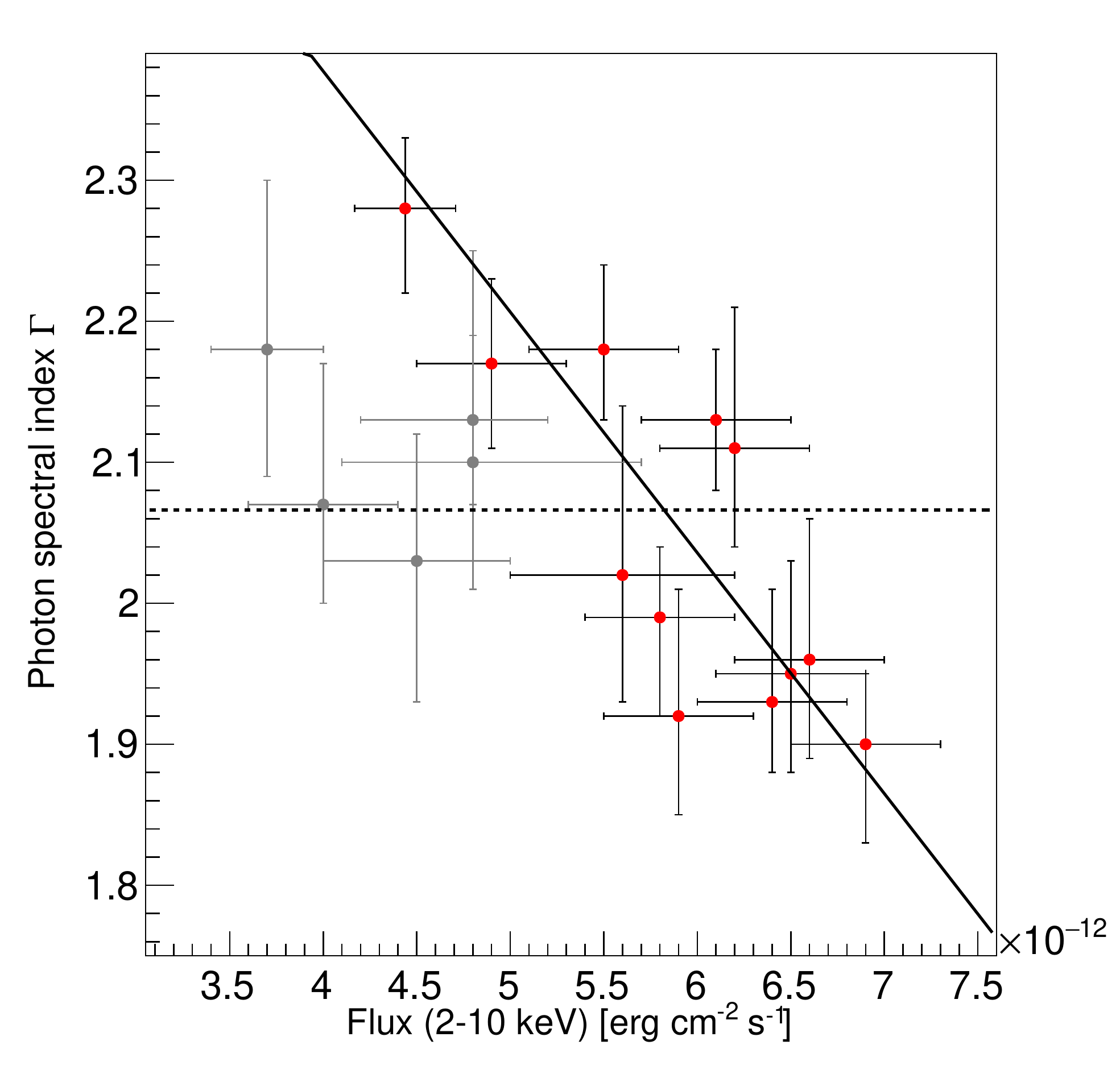}
    \caption{Power-law index of X-ray spectra as a function of the energy flux (2--10\,keV) measured by \textit{Swift}-XRT. The dashed and solid lines represent a fit to the data with a constant and a linear function, respectively. Red points show the results from individual pointing during the MWL campaign. Additionally, gray points represent the 2012 January data.}
    \label{XRay2}
\end{figure}

Comparing the light curve with the temporal evolution of the photon index yields evidence for a spectral hardening with increasing energy flux.
We fit the photon index versus time with a constant, yielding $\chi^2$/d.o.f.$=45.2/11$ (probability of $4.5\times10^{-6}$) indicating a change of the photon index from day to day. 
This evidence is also found when displaying the photon index as a function of the energy flux between 2--10\,keV as shown in Fig.~\ref{XRay2}.
A linear fit gives a $\chi^2$/d.o.f. of $14.0/10$ (probability of 0.17). Thus, a harder-when-brighter behaviour is observed during the campaign. Although spectral and flux variability in the X-ray band were previously reported in Aleksi{\'c} et al. (\cite{aleksic14a}), a harder-when-brighter trend for these observations was not found.

The hydrogen column density stayed constant during the campaign ($\chi^2$/d.o.f.$=13.4/11$, probability of 0.26 for a constant fit) and is consistent 
with the Galactic value for IC\,310 ($0.12\times10^{22}$\,cm$^{-2}$) from Kalberla et al. (\cite{kalberla10}) taking into account the large systematic uncertainties ($\sim30\%$) of the survey. Hence, no conclusions can be drawn on intrinsic photo-absorption during the campaign.

\begin{table*}
\small
\caption{\textit{Swift}-XRT X-ray spectral measurements from observations in 2012 January and 2012 November to December. Energy fluxes in the range 2-10 keV were determined by a simple absorbed power-law fit. The photon index $\Gamma$ is defined following $F\propto E^{-\Gamma}$. $N_\mathrm{H}$ denotes the absorption with an equivalent column of hydrogen.}             
\label{table_LC_swift}     
\centering       
\renewcommand{\arraystretch}{1.5}
\begin{tabular}{c c c c c c c}       
\hline\hline
   Obs. ID  &MJD start     & Exps.  & $F_{2-10\,\mathrm{keV}}\times10^{-11}$ & $\Gamma$ & $N_{\mathrm{H}}$ & $\chi^2$/d.o.f.\\    
		  	   & 	    & [s]    & [erg\,s$^{-1}$\,cm$^{-2}$] & &[10$^{22}$\,cm$^{-2}$] & \\
\hline 
   2012 Jan. &              & 13214  & $0.438\pm0.015$  &$2.10^{+0.05}_{-0.04}$ & $0.125^{+0.015}_{-0.029}$&  58.9/75               \\
\hline 
00032264001 & 55952.65 & 2989 & $0.40\pm0.04$          & $2.07^{+0.10}_{-0.07}$ & $0.10^{+0.06}_{-0.04}$ & 18.9/17\\
00032264003 & 55953.58 & 2885 & $0.37\pm0.03$          & $2.18^{+0.12}_{-0.09}$ & $0.18\pm0.05$ & 11.7/15\\
00032264004 & 55954.12 & 2742 & $0.48^{+0.09}_{-0.07}$ & $2.10^{+0.15}_{-0.09}$ & $0.14\pm0.04$ & 9.3/17\\
00032264005 & 55955.12 & 3094 & $0.48^{+0.04}_{-0.06}$ & $2.13\pm0.06$          & $0.12\pm0.04$ & 24.5/22\\
00032264006 & 55956.14 & 1504 & $0.45\pm0.05$          & $2.03^{+0.09}_{-0.10}$ & $0.11^{+0.05}_{-0.06}$    & 14.6/8\\		  	   
\hline 
   2012 Nov.-Dec. &              & 45773  & $0.606\pm0.010$  &$2.036^{+0.022}_{-0.019}$ & $0.135\pm0.008$&  222.4/240               \\
\hline 
00032264007 & 56245.67 & 4977 & $0.64\pm0.04$ & $1.93^{+0.08}_{-0.05}$ & $0.126^{+0.024}_{-0.026}$ & 37.3/38\\
00032264008 & 56253.09 & 3984 & $0.61\pm0.04$ & $2.13\pm0.05$          & $0.118^{+0.035}_{-0.025}$ & 43.8/36\\
00032264009 & 56253.98 & 3968 & $0.69\pm0.04$ & $1.90^{+0.05}_{-0.07}$ & $0.061^{+0.024}_{-0.025}$ & 32.9/31\\
00032264010 & 56255.70 & 3991 & $0.49\pm0.04$ & $2.17\pm0.06$          & $0.151^{+0.030}_{-0.068}$ & 22.3/23\\
00032264011 & 56273.95 & 1983 & $0.56\pm0.06$ & $2.02^{+0.12}_{-0.09}$ & $0.16^{+0.09}_{-0.06}$    & 7.3/11\\
00032264012 & 56275.88 & 3891 & $0.59\pm0.04$ & $1.92^{+0.09}_{-0.07}$ & $0.116^{+0.029}_{-0.025}$ & 29.6/28\\
00032264013 & 56276.81 & 3878 & $0.66\pm0.04$ & $1.96^{+0.10}_{-0.07}$ & $0.140^{+0.027}_{-0.024}$ & 34.3/31\\
00032264014 & 56277.81 & 3878 & $0.65\pm0.04$ & $1.95^{+0.08}_{-0.07}$ & $0.13^{+0.04}_{-0.05}$    & 23.2/30\\
00032264015 & 56278.82 & 3660 & $0.62\pm0.04$ & $2.11^{+0.10}_{-0.07}$ & $0.171^{+0.029}_{-0.031}$ & 37.7/29\\
00032264016 & 56279.82 & 3864 & $0.55\pm0.04$ & $2.18^{+0.06}_{-0.05}$ & $0.173^{+0.027}_{-0.030}$ & 23.1/30\\
00032264017 & 56280.82 & 3856 & $0.44\pm0.03$ & $2.28^{+0.05}_{-0.06}$ & $0.149\pm0.028$           & 47.3/28\\
00032264018 & 56281.82 & 3844 & $0.58\pm0.04$ & $1.99^{+0.05}_{-0.07}$ & $0.127^{+0.029}_{-0.024}$ & 31.9/28\\
\hline 
\end{tabular} 
\end{table*}

\begin{figure}
    \centering
    \includegraphics[width=8cm]{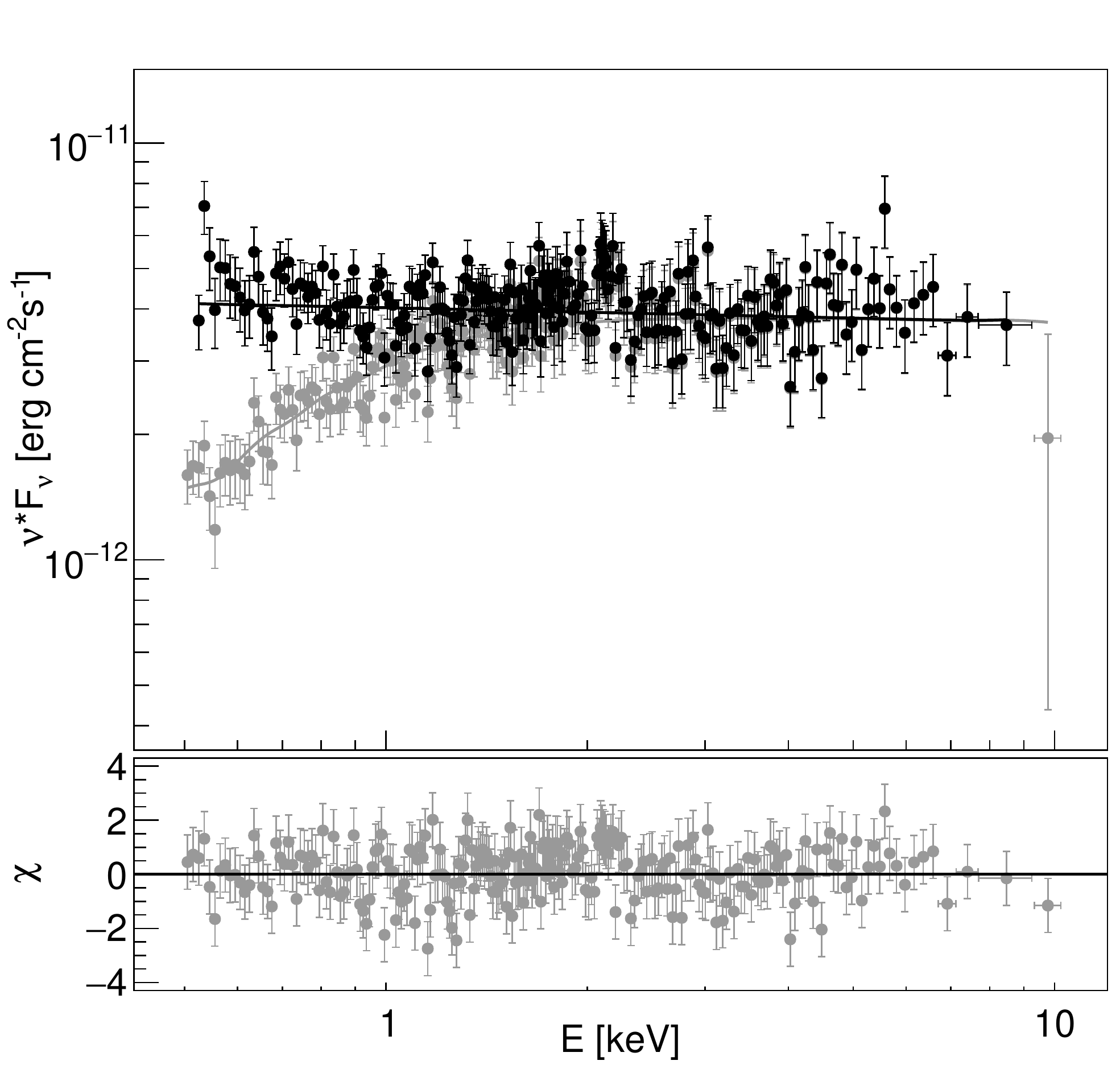}
    \caption{Averaged spectral energy distribution from \textit{Swift}-XRT observations from 2012 November to December. The measured spectrum is shown with gray points and is fitted with an absorbed power-law function (gray lines). Fit results are given in Table~\ref{table_LC_swift}. The de-absorbed spectrum is shown with black points and black lines. \textit{Top panel:} Resulting spectra in the energy range 0.5--10\,keV. \textit{Bottom panel:} Residual of the observed spectrum. }
    \label{XRay3}	
    \end{figure}

For the investigation of the multi-wavelength SED, all data taken in 2012 November and December were combined to derive an averaged spectrum during the campaign. The result is shown in Fig.~\ref{XRay3}. It can be well described with a simple absorbed power-law in the range 0.5--10\,keV with a hard photon spectral index of $\Gamma=2.036^{+0.022}_{-0.019}$. This index is comparable within the errors with the index of the averaged spectrum obtained from measurements at the beginning of the year 2012.

From the \textit{INTEGRAL} data, a 1\,$\sigma$ upper limit is extracted from the variance of the stacked mosaic image at the source position taking into account a photon index of 2.0, which is extrapolated from the \textit{Swift}-XRT band. This results in an upper limit of the energy flux of $2.9\times10^{-12}$\,erg\,cm$^{-2}$\,s$^{-1}$ at 110\,keV in the energy range 20--200\,keV. Assuming a softer photon index of 2.5, the upper limit yields $1.7\times10^{-12}$\,erg\,cm$^{-2}$\,s$^{-1}$ at 110\,keV in the energy range 20--200\,keV. For the broad band SED in Fig.~\ref{MWL_SED}, we show the 2\,$\sigma$ upper limit calculated for the photon index of 2.0 and multiply it with a factor of 1.5 for the root mean square of the significance map.

The BAT data yield an energy flux of $(3.2\pm1.4)\times10^{-12}$\,erg\,cm$^{-2}$\,s$^{-1}$ for the energy range 20--100\,keV and a photon index of $1.8\pm1.1$ for the 104-month \textit{Swift}-BAT survey maps with a signal-to-noise ratio of 1.85\,$\sigma$. Uncertainties are given at a 90\% confidence level. For the broad-band SED, we use the energy flux value instead of an upper limit. We apply the criterion of $3\,\sigma$ for the calculation of energy flux upper limits. But, as three times the $1\,\sigma$ energy flux uncertainty is smaller than the energy flux measurement, we do not classify the energy flux as an upper limit.

    
\subsection{Optical and UV}

\begin{figure}
    \centering
      \includegraphics[width=9.cm]{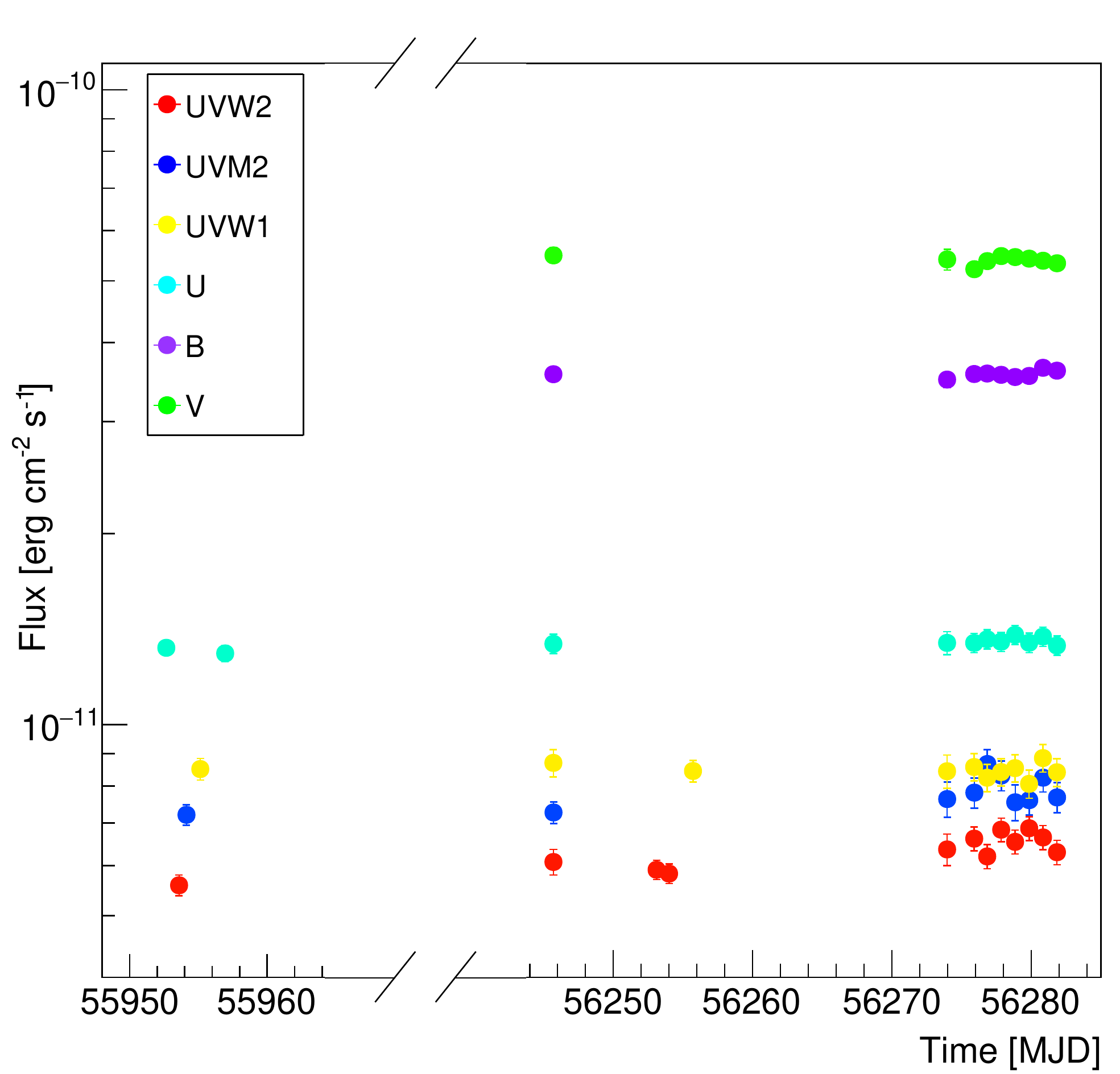}
      \caption{Multi-frequency energy flux light curve of IC\,310 obtained from \textit{Swift}-UVOT observations. The markers for the different frequencies are indicated in the legend. All data are de-reddened using $N_\mathrm{H}=0.13\times10^{22}\,\mathrm{cm}^{-2}$. To show the results from the 2012 January as well as the 2012 November to December observations simultaneously, the x-axis has been interrupted, as indicated by the diagonal lines. Note that the y-axis uses a logarithmic scale.  }
               \label{UVOT_lightcurve}%
 \end{figure}

The KVA R-band optical light curve is shown in Fig.~\ref{MWL_lightcurve} and does not show any signs of variability. 
The light curve in the time range of 2012 November 12 to 2013 February 02 (MJD\,56243--56325) 
 is consistent with a constant flux density of $(9.08\pm0.04)$\,mJy ($\chi^2$/d.o.f.$=5.3/17$, 
probability of $0.99$).
As no further historical monitoring in the R-band was conducted for IC\,310, the flux density cannot be compared with other measurements.\\
\begin{table}
\caption{\textit{Swift}-UVOT observations.}            
\label{table:UVOT_1}      
\centering                         
\begin{tabular}{c c c c c}        
\hline\hline
Obs. ID    	& MJD start & Exps. & Filter \\
000...		&  	    & [ks]   &		\\
\hline
32264001	& 55952.65	    & 2.9 	  &U	\\
32264003	& 55953.58	    & 2.8	  &UVW2	 \\
32264004	& 55954.12	    & 2.7	  &UVM2	\\
32264005	& 55955.12	    & 3.0	  &UVW1	 \\
32264006	& 55956.14	    & 1.0	  &U	 \\
\hline
32264007	& 56245.67	    & 4.8	  &all	\\
32264008	& 56253.09	    & 3.9         &UVW2	\\
32264009	& 56253.98	    & 3.9	  &UVW2	 \\
32264010	& 56255.70	    & 3.9	  &UVW1	 \\
32264011	& 56273.95	    & 1.9	  &all	 \\
32264012	& 56275.88	    & 3.8	  &all	 \\
32264013	& 56276.81	    & 3.7         &all	 \\
32264014	& 56277.81	    & 3.7	  &all	 \\
32264015	& 56278.82	    & 3.5	  &all	\\
32264016	& 56279.82	    & 3.7	  &all	 \\
32264017	& 56280.82	    & 3.7	  &all	 \\
32264018	& 56281.82	    & 3.7	  &all	 \\
\hline      
\end{tabular}
\end{table}
An overview of the observations from 2012 November to December by \textit{Swift}-UVOT is given in Table~\ref{table:UVOT_1} together with the 2012 January measurements.   
The results show no significant variability for any of the filters (Fig.~\ref{UVOT_lightcurve}). A constant fit to the 2012 November to December data revealed no variability for B, U, V, UVW1 with a fit probability of 0.98, 0.99, 0.95, and 0.98, respectively. Smaller fit probabilities were obtained for the UVM2 and UVW2 measurements (0.26 and 0.03). Comparing the measurements in 2012 January and 2012 November-December, only at the highest frequencies (UVM2 and UVW2) the mean flux densities from late 2012 are not compatible within the errors with the measurements in early 2012.  \\   
To interpret and model the broad band SED, mean flux densities for each frequency were calculated from the data taken in 2012 November to December. The values obtained are listed in Table~\ref{table:UVOT_2}. 
\begin{table}
\caption{\textit{Swift}-UVOT optical/UV flux density measurements from observations in 2012 November to December. The third and fourth column report the observed and de-reddened mean values, respectively.}            
\label{table:UVOT_2}      
\centering                         
\begin{tabular}{c c c c}        
\hline\hline
 Filter & Freq.	& $F\times 10^{-12}$         & $F_{\mathrm{de-reddened}}\times 10^{-12}$\\
        & [Hz]	& [erg\,cm$^{-2}$\,s$^{-1}$] & [erg\,cm$^{-2}$\,s$^{-1}$]\\
\hline
V      &$5.48\times10^{14}$	& $27.90\pm1.18$ & $53.96\pm2.28$ \\
B      &$6.83\times10^{14}$	& $14.82\pm0.54$ & $35.64\pm1.29$\\
U      &$8.65\times10^{14}$	& $4.69\pm0.25$& $13.58\pm0.0.73$\\
UVW1     &$1.15\times10^{15}$	& $2.05\pm0.16$& $8.52\pm0.67$\\
UVM2     &$1.33\times10^{15}$	& $1.09\pm0.09$& $7.86\pm0.66$\\
UVW2     &$1.55\times10^{15}$	& $1.10\pm0.08$& $6.35\pm0.46$\\
\hline      
\end{tabular}
\end{table}
The SED in Fig.~\ref{MWL_SED} shows the observed and de-reddened KVA and mean data from \textit{Swift}-UVOT using the  absorption measurement from \textit{Swift}-XRT. 
\subsection{Radio band}

The light curve at 15\,GHz obtained by the OVRO monitoring program is shown in Fig.~\ref{MWL_lightcurve}.
From a fit of the data points from 2012 October 30 to December 23 (MJD\,56230-56284) to a constant, a mean flux density of $(0.151\pm0.002)$\,mJy is measured with a $\chi^2$/d.o.f. of $31.1/13$ and rather low probability of $0.003$ for being constant.

From the EVN and MOJAVE data, the core and total flux density were calculated and included in Fig.~\ref{MWL_SED}. The estimated uncertainties for the EVN flux density measurements at 1.7, 5.0, 8.4\,GHz are 10\% and for 22\,GHz, 15\%. For the MOJAVE data we adopt an uncertainty of 5\% for the total flux density \cite{lister2005}.
We use the results from the 5.0\,GHz observation taken from Aleksi{\'c} et al. (\cite{aleksic14b}).

\section{Discussion}

In the following we discuss the results in terms of multi-frequency flux variability and interpret the broad-band SED.

\subsection{Multi-frequency flux variability}

\begin{figure*}
    \centering
      \includegraphics[width=14cm]{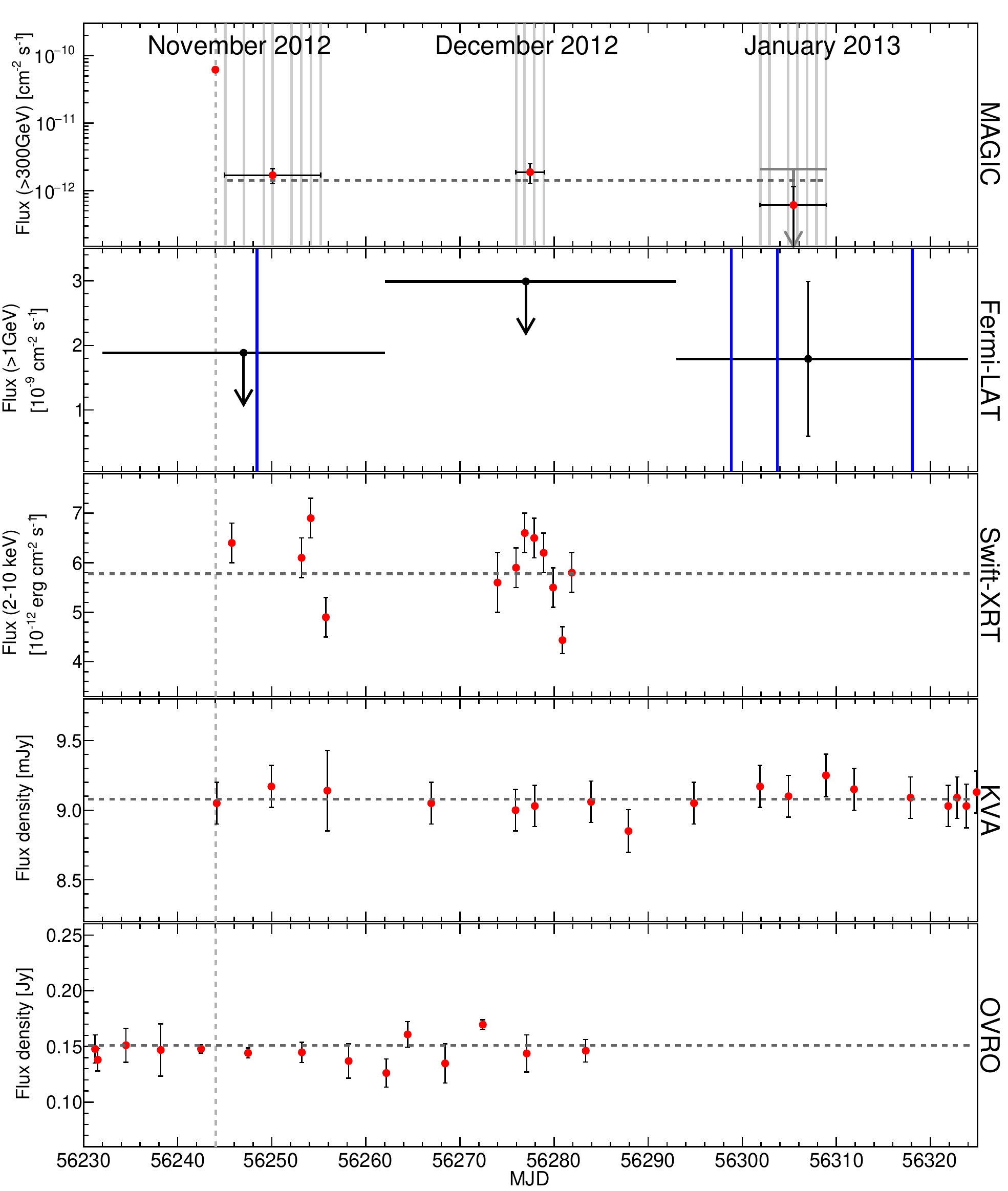}
      \caption{Multi-wavelength light curve of IC\,310 between 2012 November and 2013 January. The vertical gray dashed line indicates the day of the TeV flare. The horizontal gray dashed lines are the fits to the data with a constant. \textit{From top to bottom:} MAGIC monthly flux measurements above 300\,GeV. In addition, for 2013 January a flux 95\% confidence level upper limit is calculated. The gray lines indicate the individual days when observations were performed with MAGIC. Note that the y-axis uses a logarithmic scale.  \textit{Fermi}-LAT above 1\,GeV, blue lines indicate the arrival times of gamma-ray event candidates. \textit{Swift}-XRT fluxes between 2--10\,keV. R-band flux measurements by KVA (not corrected for the contribution of the host galaxy). OVRO fluxes at 15 GHz. }
               \label{MWL_lightcurve}%
 \end{figure*}

The combined multi-wavelength light curve is shown in Fig.~\ref{MWL_lightcurve}. An exceptional TeV flare was found by MAGIC in November 2012. IC\,310 remained active at VHE even after the flare. Only \textit{Fermi}-LAT, \textit{Swift}-BAT and partially KVA observed simultaneous to MAGIC on 2012 November 12-13. However, IC\,310 could not be detected with the first two instruments during the TeV flare and the optical emission observed by KVA is dominated by the host galaxy.

Contemporaneous measurements starting after the flare indicated a high and variable state of the object in the soft X-ray range. This marks the first time that a contemporaneous measurement of the VHE and the soft X-ray flux is reported for IC\,310. 

Conclusions on a correlation between the two bands should be drawn with caution because historical simultaneous measurements in both energy ranges are missing. However, considering the harder-when-brighter trend in the VHE and X-ray bands reported in Fig.~\ref{VHESpectraCom} and Fig.~\ref{XRay2}, one can speculate that these two bands may be connected.

A harder-when-brighter behaviour is frequently observed for other sources such as high-frequency peaked BL\,Lac (HBL) objects (\cite{pian98}; \cite{giommi00}; \cite{aleksic13}; \cite{furniss15}; \cite{aleksic15}; \cite{kapanadze16}; \cite{ahnen16}; \cite{balokovic2016}). Here, the increase of the X-ray and VHE flux is combined with a hardening of the spectrum. This is consistent with the synchrotron self-Compton (SSC) mechanism, see Sect. 4.2.2. The first hump in the SED moves towards higher frequencies. The same happens for the second hump, as during a higher flux state the synchrotron photons in the low energy band are seeds for the inverse-Compton emission observed in the high-energy bands. 
If such a behaviour is observed for a blazar with a small viewing angle, one should expect the same also for radio galaxies. This would be in-line with the unified model by Urry \& Padovani (\cite{urry95}). Note that X-ray data from other misaligned blazars (e.g., NGC\,1275) often show complex emission, with a non-pure power-law and extended jet emission or X-ray emission of the filaments \cite{fabian2011}.  

The optical light curve does not show a significantly higher flux or variability. The lack of optical variability is consistent with the result of the SED modeling (see Sect. 4.2.3). The optical-infrared emission is ascribed entirely to the host galaxy, while the variable jet emission starts to dominate in the far-UV and soft X-ray band. The radio light curve indicates weak variability. As radio flares combined with an appearance of a new radio component in the VLBI images are sometimes found a few months after a gamma-ray flare \cite{acciari09}, the time period covered here is too short to draw conclusions. 

Generally, the VHE flare might be interpreted as an injection of fresh electrons and positrons into the SSC emission region. One possible origin of these particles could be the electro-magnetic cascades in the gap region of magnetospheric models resulting in an increased number of $e^+e^-$-pairs (\cite{beskin92}; \cite{rieger00}; \cite{neronov07}; Levinson \& Rieger \cite{levinson}). Such a model has been used to explain the flaring activity from M\,87 by Levinson \& Rieger (\cite{levinson}) as well as the minute time-scale variability of the gamma-ray emission of IC\,310 at the beginning of the MWL campaign reported by Aleksi{\'c} et al. (\cite{aleksic14b}) and critically examined in \cite{hirotani16}. Since the magnetic field is assumed to be along the jet axis, these particles should then move along the jet and maybe also into the emission region which we observed after the flare. This could cause a higher flux in the X-ray regime as well as at VHE. Note that the quiescent state of IC\,310 might be undetectable with MAGIC as sometimes IC\,310 could not be detected over a long 
observation time (\cite{aleksic12}). 
Furthermore, these fresh particles moving along the jet could also explain a general activity in the radio band. 

\subsection{Modeling of the Spectral Energy Distribution}

\begin{figure*}
    \centering
      \includegraphics[width=14cm]{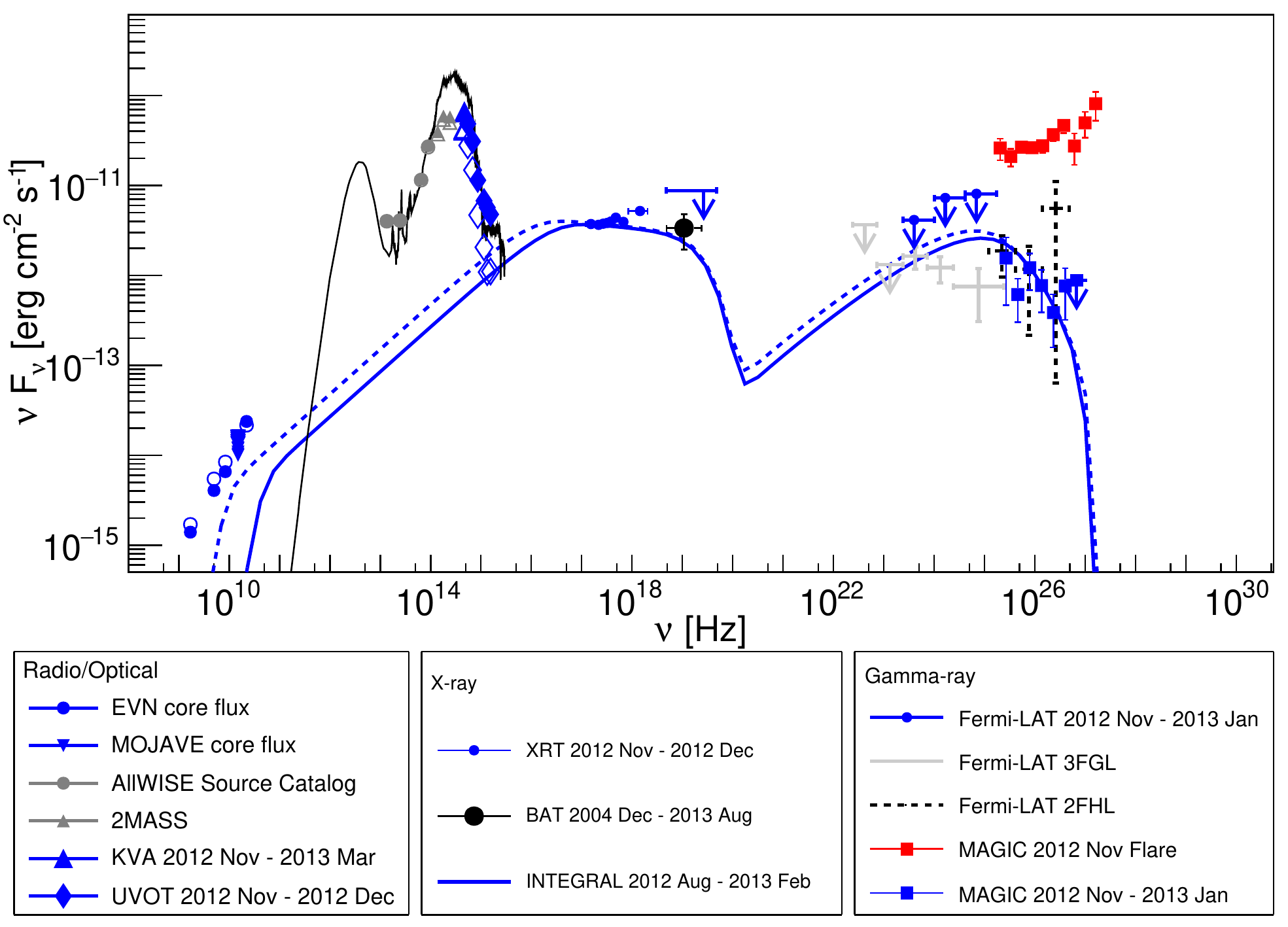}
      \caption{Contemporaneous broad-band spectral energy distribution during the MWL campaign from 2012 November to 2013 January (blue points). Individual data points are explained in the legends. Long-term averaged (if they outlast the campaign) and archival measurements are shown in black and gray, respectively. The MAGIC data points were de-absorbed using the EBL model by \cite{dominguez11}. The red points show the derived SED of the flare observed at the beginning of the campaign, and reported in Aleksi{\'c} et al. (\cite{aleksic14b}). The \textit{Fermi}-LAT upper limits were calculated for the time period 2012 November to 2013 January. 
      In addition, the data points from the 3FGL \cite{acero2015} and 2FHL catalogs \cite{ackermann16} are shown. The XRT data were averaged over the entire campaign, de-absorbed with an equivalent column of hydrogen and re-binned to eight points. The data from 2MASS, AllWISE, KVA, and UVOT (open markers) have been de-reddened (filled markers) according 
to the XRT data. The open markers in the radio band indicate the measured total flux density, whereas the filled markers are the core flux density values. The 5.0\,GHz measurement is taken from Aleksi{\'c} et al. (\cite{aleksic14b}). The blue lines were obtained from a modeling of the data with a single-zone SSC model from \cite{krawczynski04}. The solid/dashed line is the model with a viewing angle of 10$^\circ$/20$^\circ$, respectively. The black line represents a template SED for an S0 host galaxy from Polletta et al. (\cite{Polletta07}). }
    \label{MWL_SED}%
 \end{figure*}

\subsubsection{Simultaneity of the data}

Due to the variable behaviour of IC\,310 in some frequency bands (Aleksi{\'c} et al. \cite{aleksic14a}, \cite{aleksic14b}), SED fitting requires simultaneous data. In the VHE regime, the emission of IC\,310 after the flare was rather faint, but still detectable over a long time range spanning over three months. Therefore, we calculate one spectrum over the entire observation time. For the SED, one averaged spectrum from the \textit{Swift}-XRT observation was calculated. In the high energy range, the \textit{Fermi}-LAT measurements are covering the time period from 2012 November 1 to 2013 January 31. The HE data is contemporaneous to the MAGIC and \textit{Swift}-XRT/UVOT observations. Furthermore, we include the archival spectra from the 3FGL \cite{acero2015} and 2FHL catalogs \cite{ackermann16}. The former was recorded from 2008 August - 2012 July and does not cover the time of the campaign, whereas the 2FHL outlasts the campaign. The results from BAT were extracted from the long-term 104-month survey running from 
2004 December to 2013 August, and are taken as an average measurement. \textit{INTEGRAL} observations cover the time range from 2012 
August to 2013 February and therefore are quasi-simultaneous.  The optical data from KVA were 
coordinated with the MAGIC observations and are therefore simultaneous to those. The radio data from EVN and MOJAVE have been included on a quasi-simultaneous basis. To model the host galaxy, we also include non-simultaneous data from 2MASS and \textit{WISE} which were recorded during 1997 June to 2001 February and 2010 January to 2010 November, respectively. 
 
\subsubsection{Model description}

\begin{table*}
\caption{Model parameters for the one-zone SSC model describing the broad-band SED of IC\,310. The resulting model curves are depicted in Fig.~\ref{MWL_SED}.}            
\label{table:modelparameter}      
\centering                         
\begin{tabular}{c c c c c c c c c c c c}        
\hline\hline
$\Gamma$ & $\delta$ & $\theta$ & $p_1$ & $p_2$ & $E_\mathrm{min}$ & $E_\mathrm{max}$ & $E_\mathrm{break}$ & $U_e$ & $\eta$ & $B$ & $R$ \\
	 &	    & [$^\circ$]&      &	&log10($E$/eV)	 & log10($E$/eV)	    & log10($E$/eV)        & [erg/cm$^{3}$] & [$U_e/U_B$]     & [G] & [cm] \\
\hline
5	 &5.7 	    & 10       & 2.0   &3.1	 &  5.7         & 12.0                  & 10.6              & 0.21 & 59  & 0.3 &  $2.1\times10^{15}$      \\   
5	 &2.5 	    & 20       & 2.0   &3.1	 &  5.7         & 12.4                  & 10.8              & 0.012 & 30  & 0.1 &  $3.0\times10^{16}$      \\
\hline      
\end{tabular}
\end{table*}

The SED of a jet-dominated radio-loud AGN can be explained by non-thermal emission of accelerated particles. In the shock-in-jet model \cite{blandford79}, these particles are accelerated at shock waves in the jet which can originate, e.g., from density fluctuations of the plasma. The low-energy emission of the double-humped SED of radio-loud AGNs is explained by synchrotron radiation of the electrons and positrons due to the magnetic fields in the jets. In contrast, the origin of the higher energy hump is still a matter of debate. Leptonic- i.e., inverse-Compton scattering (\cite{marscher85}; \cite{maraschi92}; \cite{bloom96})- and hadronic- e.g., proton synchrotron (\cite{mannheim93a}; \cite{muecke03}) and neutral pion decay initiated electromagnetic cascades \cite{mannheim93b}- processes can lead to the emission observed at higher frequencies. As shown in Aleksi{\'c} et al. (\cite{aleksic14b}), this shock-in-jet model cannot explain the fast TeV flare in 2012 November. 

As mentioned in Sect.~4.1, freshly injected electrons and positrons may result from electro-magnetic cascades inside a gap region of a pulsar-like magnetosphere surrounding the central black hole. The acceleration of the particles due to the huge electric potential in the gap and emission via inverse-Compton and curvature radiation can explain the flaring behaviour as reported in Aleksi{\'c} et al. (\cite{aleksic14b}).  However, after some time, the gap short-circuited because the charge density reaches the Goldreich-Julian charge density and the cascading stops. The particles may move in the direction of the jet. 

We adopt the single-zone SSC model by \cite{krawczynski04} to model the broad-band SED. The model consists of the following parameters: the bulk Lorentz factor $\Gamma_\mathrm{b}$, the viewing angle $\theta$, the magnetic field $B$, the radius of the emission region $R$, the electron energy density $U_e$, the ratio $\eta$ of the electron energy density to the magnetic field energy density $U_B$, the minimal energy and the maximal energy of the electrons $E_\mathrm{min}$ and  $E_\mathrm{max}$, the break energy $E_\mathrm{break}$, and the spectral indices $p_i$ ($dN/dE\propto E^{-p_i}$, $E$ is the electron energy in the jet frame). In this model it is assumed that the electrons follow a broken power-law energy spectrum with the indices $p_1$ for the energies $E_\mathrm{min}$ to $E_\mathrm{break}$ and $p_2$ for $E_\mathrm{break}$ to $E_\mathrm{max}$.

We assume a viewing angle of $10^\circ\lesssim\theta\lesssim20^\circ$ as found by Aleksi{\'c} et al. (\cite{aleksic14b}). For this range of viewing angles, a rather low bulk Lorentz factor is necessary to enable having at least a moderate Doppler boosting. Here, we use $\Gamma_\mathrm{b}=5$. Note that the models applied to the data are only two possibilities, one for $\theta=10^\circ$ and one for $\theta=20^\circ$; generally, a large degeneracy of parameters exists \cite{ahnen2016}. 

In addition to the SSC model, a template SED for a galaxy of morphology type S0 (Polletta et al. \cite{Polletta07}) is added to account for the strong dominance of the host galaxy in the infrared to UV range as observed from the S0-type galaxy IC\,310 \cite{deVaucouleurs91}.\footnote{The template from the \textit{Spitzer} Wide-area InfraRed Extragalactic survey (SWIRE) library can be found under the following link: \url{http://www.iasf-milano.inaf.it/~polletta/templates/swire_templates.html}}  

\subsubsection{Results}

The broad band SED is shown in Fig.~\ref{MWL_SED}. Apart from the radio telescopes, the angular resolution of all other instruments used in this paper is a limitation to study a jetted object such as IC\,310. Detailed radio observations do not indicate, e.g., lobe emission (\cite{sijbring98}; \cite{kadler12}). Moreover, deep X-ray measurements show a mostly point-like emission positionally consistent with the bright radio core (Dunn et al. \cite{dunn10}). Therefore, when modeling the broad-band SED, we consider that all emission is coming from around the central region of the AGN.

The core fluxes from the radio VLBI data follow a typical flat spectrum with a constant flux density. The hard spectrum in the soft X-ray regime suggests that the synchrotron emission peaks in the $\nu F_\nu$ representation in the soft X-ray band and declines in the hard X-ray band as indicated by the \textit{Swift}-BAT and \textit{INTEGRAL} data, though both are not strictly simultaneous to the \textit{Swift}-XRT measurements. The \textit{Fermi}-LAT upper limits are consistent with the long-term spectrum from the 3FGL catalog. The MAGIC spectrum of the campaign yields a similar flux state as the 2FHL catalog data and also follows the high energy points of the 3FGL spectrum. These three spectra, even though they only have a partial temporal overlap, show that the high energy spectral index is very hard in the GeV range ($\Gamma_\mathrm{3FGL}=1.90\pm0.14$, $\Gamma_\mathrm{2FHL}=1.34\pm0.42$) and that the second hump is located below 100\,GeV during the campaign. However, during the TeV flare, the peak 
frequency of the second hump is above 10\,TeV and therefore, shifted by more than two orders of magnitude. The hard VHE spectrum extending to 10\,TeV cannot be reproduced within a one-zone SSC scenario because of the strong Klein-Nishina suppression at multi TeV energies. Instead, \cite{hirotani16} could reproduce the flux and the hard spectrum during the flare of IC\,310 with a superposition of curvature emission in the magnetosphere with varying curvature radii. Spectral variability in the TeV band can be explained by, e.g., different accretion rates in this model.

In general, the SED reveals that the object has been measured in a state of rather low TeV activity after the VHE flare on 2012 November 12-13 and in a high state in the X-ray band. This is the first time that both regimes were measured simultaneously, so one cannot make a statement concerning the occurrence of a simultaneous low and high state in both bands. 

The measured SED of IC\,310 seems to follow a simple, featureless double-hump structure besides the host galaxy emission, as seen from blazars. 
A single-zone SSC model is able to explain the broad-band emission in general, though it is not possible to reproduce the flux densities of the VLBI core measurements which likely comprise additional radio emission from other regions.  The radio and X-ray data suggest a broad synchrotron emission hump which can be explained with a large maximal Lorentz factor of the particles. This agrees with the second hump peaking in the GeV range or even higher. However, it is difficult to fit the 3FGL, 2FHL, and low-state MAGIC data simultaneously with an SSC model due to the broadness of the second hump. The SSC model curves are, however, consistent with the upper limits in the GeV band calculated for this campaign.

The parameters of the SSC models used for the magnetic field, the radius of the emission region, and the necessary Lorentz factors of the particles are in agreement with the constraints from the magnetospheric model used to explain the emission from the TeV flare (Aleksi{\'c} et al. \cite{aleksic14b}). Here, we assume that the emission region is caused by a cloud formed out of the particles created due to the cascading in the gap region during the flaring period. This blob moves away from the center of the AGN in the direction of a conical jet with a distance $d$ corresponding to a travel time of 1-2 months. Here, the magnetic field strength reduces from the values inferred from Levinson \& Rieger (\cite{levinson}) to the values used in the SSC models from this work, assuming the dependence of $B(\mathrm{d})\sim d^{-1}$ (\cite{koenigl1981}; \cite{lobanov1998}). Furthermore, assuming a conical geometry, the radius of the emission region increases to $R\sim10^{15}$\,cm, consistent with the values used for the $10^\circ$ model.\footnote{We assumed a radius at the 
starting point consistent with the variability time scale observed in Aleksi{\'c} et al. (\cite{aleksic14b}).} The required energetics of maximal $E_\mathrm{max}=12$ (in units of $\mathrm{log10}(E/\mathrm{eV})$) with a Lorentz factor of the particles of $\sim 10^6$ can be achieved, as even higher values were found by \cite{hirotani16}.          
The SSC code of \cite{krawczynski04} includes the effect of $\gamma\gamma$-pair production and for both models the optical thickness for $\gamma\gamma$-pair production is low enough for TeV emission to escape.

In the IR-optical-UV range of the SED, the emission is strongly dominated by the host galaxy. The template SED of the S0-type galaxy represents the measured data points very well. An SED for an elliptical galaxy would have overestimated the contribution in the optical range by one order of magnitude. Instead, the template we used suggests a rather strong emission in the IR band due to starburst processes. Unfortunately, NGC\,1275 is too bright in the far-infrared and microwave regime, so that the emission from IC\,310 could be outshone by NGC\,1275 in maps obtained by the \textit{Planck} satellite.\footnote{However, the angular resolution of \textit{Planck} at high frequencies should be good enough to separate the emission from both AGNs.} 

Finally, we use the broad-band SED to determine the mass of the central black hole analogous to \cite{Krauss16} using the fundamental plane of black hole activity. We assume the X-ray flux of the 2012 November and December data (Table~\ref{table_LC_swift}), the 5\,GHz core flux measurement from EVN and extrapolated the EVN core data with a simple power-law spectrum in order to estimate the 1.4\,GHz flux. The extrapolation was done by fitting the spectrum of the 1.7, 5.0, 8.4, and 22.2\,GHz core data with a power-law of the form $\mathrm{d}N/\mathrm{d}E=(E/E_0)^{-\Gamma_\mathrm{R}}$ with $E_0$ being a normalization flux and $\Gamma_\mathrm{R}$ the power-law index. The fitting yields $\Gamma_\mathrm{R}=0.95$ and thus a differential flux at 1.4\,GHz of $1.08\times10^{-15}$erg\,cm$^{-2}$\,s$^{-1}$.
The resulting masses are given in $\mathrm{log}_{10}(\mathrm{M}_{\odot})$: $(7.89\pm0.09)$ for the fundamental plane found in \cite{2009ApJ...706..404G},  $(8.4\pm1.3)$ for \cite{2003MNRAS.345.1057M}, $(9\pm4)$ for
\cite{2006A&A...456..439K}, for $(8.4\pm2.9)$ \cite{2013MNRAS.429.1970B}, and $(8.2\pm1.0)$ for \cite{2016MNRAS.455.2551N}.\footnote{The values based on Eq 4. and the parameters from Eq. 6 by \cite{2009ApJ...706..404G} seem to under-predict the uncertainties of the black hole mass estimate.} 
These values are consistent with $M_{\mathrm{BH}}\simeq\left(3^{+4}_{-2}\right)\times\,10^{8}$\,M$_{\odot}$ reported in Aleksi{\'c} et al. (\cite{aleksic14b}).  Instead, \cite{berton2015} estimated a mass for IC\,310 of one order of magnitude lower than the value estimated in Aleksi{\'c} et al. (\cite{aleksic14b}) based on a smaller velocity dispersion used for the $M_{\mathrm{BH}}-\sigma$ relation. Assuming a smaller mass would somewhat weaken the arguments given in Aleksi{\'c} et al. (\cite{aleksic14b}) as the variability time scale then roughly equals the event horizon light crossing time. But the opacity problem as explained in Aleksi{\'c} et al. (\cite{aleksic14b}) still remains. 

\subsubsection{Comparsion with misaligned blazars and blazars}

The SSC model parameter values reported in this study for IC\,310 are similar to those obtained for misaligned blazars, e.g., PKS\,0625$-$354, 3C\,78 in \cite{fukazawa2015}, Cen\,A core in Abdo et al. (\cite{abdo2010a}), M\,87 in Abdo et al. (\cite{abdo2009b}), NGC\,1275 in \cite{abdo2009a} and Aleksi{\'c} et al. (\cite{aleksic14c}), and NGC\,6251 in \cite{migliori2011}.\footnote{Note that the broad-band SED of Cen\,A cannot be described by a simple one-zone SSC model due to the unusual gamma-ray spectrum (Abdo et al. \cite{abdo2010a}).} Compared to BL Lac objects, the values for the bulk Lorentz and Doppler factor are lower for IC\,310 as well as the other misaligned blazars. This is understandable because the Doppler boosting is weaker due to the larger viewing angle. However, to maintain some boosting at larger viewing angles, a smaller bulk Lorentz factor is mandatory. Lower bulk Lorentz factors were also found for gamma-ray blazars in a quiescent state \cite{ahnen2016}, but large Lorentz factors are needed to explain the fast 
variability observed from blazars \cite{begelman2008}. The radii for the emission region used for IC\,310 coincide with the observed variability of the object 
in the X-ray range after the TeV flare and are comparable to those used for other misaligned blazars. 
Interestingly, the models for IC\,310 show a higher  $\gamma_\mathrm{break}=10^{E_\mathrm{break}}/\mathrm{m}_\mathrm{e}\mathrm{c}^2$ similar to the SEDs from PKS\,0625$-$354, 3C\,78, and NGC\,6251, see \cite{fukazawa2015} and \cite{migliori2011}. Such high $\gamma_\mathrm{break}$ leads to higher synchrotron peak frequencies ($\sim10^{17}$\,Hz). This is in agreement with the hard spectrum in the soft X-ray and gamma-ray band observed from IC\,310. Among the misaligned blazars, IC\,310 shows one of the hardest gamma-ray spectra,  $\Gamma_{\mathrm{3FGL}}=1.90\pm0.14$ in the 3FGL \cite{acero2015}, the hardest in the 2FHL catalog \cite{ackermann16} with $\Gamma_{\mathrm{2FHL}}=1.34\pm0.42$, and $\Gamma_{\mathrm{MAGIC}}=1.8-2.4$ as reported by MAGIC here and in Aleksi{\'c} et al. (\cite{aleksic10}, \cite{aleksic14a}, \cite{aleksic14b}).

\section{Summary and Conclusions}

In this paper we have presented the results from the first multi-wavelength campaign from radio to the VHE range, conducted in 2012 November and 2013 January for the misaligned blazar IC\,310. During the campaign, an exceptionally bright VHE flare was detected showing fast flux variability in one night. There is no significant data from the lower energy bands available for that night due to low statistics of all-sky monitors such as \textit{Fermi}-LAT and \textit{Swift}-BAT. 
After the flare, the TeV flux declined rapidly. Through early 2013, only a low but still detectable VHE emission was observed. Compared to previous measurements and combining 
those with the results reported in this paper, a harder-when-brighter behaviour can be inferred. The same behaviour is found for the soft X-ray emission during the campaign. The photon spectral index hardens with increasing flux for IC\,310, which we report here for the first time. The harder-when-brighter behaviour of the X-ray and VHE emission after the TeV flare is consistent with the expectations from a one-zone SSC mechanism.

Other than the variability found in the X-ray band, the multi-wavelength light curve reveals no strong variability after the TeV flare. In the GeV band, no detection with a high significance could be inferred from the \textit{Fermi}-LAT observations. The same applies to the \textit{Swift}-BAT observation in the hard X-ray band. The host galaxy dominates completely the optical emission, and hence the optical variability of this object could not be properly evaluated. For the investigation of the variability in the radio band, a larger period and dedicated VLBI monitoring at one frequency is necessary to make further conclusions on the changes of the pc-scale jet after the flare. This will be discussed in a paper by Schulz et al. (in prep.).

As previously reported, the TeV flare cannot be explained with standard shock-in-jet models. An alternative suggestion is based on magnetospheric models for AGNs.  According to these, charge-depleted regions in AGN-magnetospheres are the birthplaces of the highest energetic particles and electromagnetic cascades. In these cascades, a large number of electrons and positrons are produced which can in principle load the jet, letting a new blob of particles travel along the jet axis. In this paper, we discussed the possibility that this blob can be the emission zone as required in a single-zone SSC model. We have shown that this simple model can explain the broad-band SED of IC\,310 observed during the campaign. Furthermore, the parameters used for the SSC modeling agree with those obtained for other gamma-ray loud misaligned blazars.    

\begin{acknowledgements}

The authors thank T.~Dauser, I.~Kreykenbohm, S.~Richter, A.~Shukla, F.~Spanier, and M.~Weidinger for the support during the preparation of observational proposals or discussion.\\
We would like to thank
the Instituto de Astrof\'{\i}sica de Canarias
for the excellent working conditions
at the Observatorio del Roque de los Muchachos in La Palma.
The financial support of the German BMBF and MPG,
the Italian INFN and INAF,
the Swiss National Fund SNF,
the he ERDF under the Spanish MINECO
(FPA2015-69818-P, FPA2012-36668, FPA2015-68278-P,
FPA2015-69210-C6-2-R, FPA2015-69210-C6-4-R,
FPA2015-69210-C6-6-R, AYA2013-47447-C3-1-P,
AYA2015-71042-P, ESP2015-71662-C2-2-P, CSD2009-00064),
and the Japanese JSPS and MEXT
is gratefully acknowledged.
This work was also supported
by the Spanish Centro de Excelencia ``Severo Ochoa''
SEV-2012-0234 and SEV-2015-0548,
and Unidad de Excelencia ``Mar\'{\i}a de Maeztu'' MDM-2014-0369,
by grant 268740 of the Academy of Finland,
by the Croatian Science Foundation (HrZZ) Project 09/176
and the University of Rijeka Project 13.12.1.3.02,
by the DFG Collaborative Research Centers SFB823/C4 and SFB876/C3,
and by the Polish MNiSzW grant 745/N-HESS-MAGIC/2010/0.\\
The \textit{Fermi}-LAT Collaboration acknowledges generous ongoing support from a number of agencies and institutes that have supported both the development and the operation of the LAT as well as scientific data analysis. These include
the National Aeronautics and Space Administration and the Department of Energy in the United States, the Commissariat \`{a} l’Energie Atomique and the Centre National de la Recherche Scientifique/Institut
National de Physique Nucl\'{e}aire et de Physique des Particules in France, the Agenzia Spaziale Italiana and the Istituto Nazionale di Fisica Nucleare in Italy, the Ministry of Education, Culture, Sports, Science and Technology (MEXT),
High Energy Accelerator Research Organization (KEK) and Japan Aerospace Exploration Agency (JAXA) in Japan, and the K.A. Wallenberg Foundation, the Swedish Research Council and the Swedish National Space Board in
Sweden. Additional support for science analysis during the operations phase is gratefully acknowledged from the Istituto Nazionale di Astrofisica in Italy and the Centre National
d’Etudes Spatiales in France.\\
The European VLBI Network is a joint facility of European, Chinese, South African and other radio astronomy institutes funded by their National research councils. \\ We acknowledge support by the MP0905 action ’Black Holes in a Violent Universe’.  The research leading to these results has received funding from the European
Commission Seventh Framework Programme (FP/2007-2013) under grant agreement No. 283393 (RadioNet3).\\
This research has made use of data from the MOJAVE database that is maintained by the MOJAVE team (Lister et al.,
2009, AJ, 137, 3718). \\ 
The OVRO 40-m monitoring program is
supported in part by NASA grants NNX08AW31G, NNX11A043G and NNX14AQ89G, and NSF grants AST-0808050 
and AST-1109911.\\
E.R. acknowledges partial support by the Spanish MINECO grants AYA2012-38491-C02-01 and AYA2015-63939-C2-2-P and by the Generalitat Valenciana project PROMETEO II/2014/057. R.S. was supported by Deutsche Forschungsgemeinschaft grant WI 1860/10-1.
F. K. acknowledges funding from the European Union’s Horizon 2020 research and innovation program under grant agreement No 653477.
This research has made use of the NASA/IPAC Extragalactic Database (NED) which is operated by the Jet Propulsion Laboratory, California Institute of Technology, under contract with the National Aeronautics
and Space Administration.\\
We would like to thank the referee for helpful comments.

\end{acknowledgements}

\end{document}